\documentclass[aps,nofootinbib,preprintnumbers,showpacs,prd,
twocolumn,superscriptaddress]{revtex4-2}
\usepackage{array}
\usepackage{booktabs}
\usepackage{subfigure}
\usepackage{capt-of}
\usepackage{graphicx}
\usepackage{lipsum}
\usepackage{epstopdf}
\usepackage{amsmath}
\usepackage{bm}
\usepackage{amssymb}
\usepackage{color,xcolor}
\usepackage[bookmarks=false]{hyperref}
\hypersetup{colorlinks=true,
	citecolor=green,
	linkcolor=blue,
	urlcolor=blue,
	pdfstartview=FitH,
	bookmarksopen=true}

\begin{document}
	
	\title{Twist-3 generalized parton distributions of sea quarks at zero skewness in the light-cone quark model}
	
	\author{Xiaoyan Luan}
	\author{Zhun Lu}\email[]{zhunlu@seu.edu.cn}
	\affiliation{School of Physics, Southeast University, Nanjing 211189, China}

\begin{abstract}
We present a systematic study of twist-3 generalized parton distributions (GPDs) for $\bar{u}$ and $\bar{d}$ sea quarks in the proton at zero skewness ($\xi=0$) using the light-cone formalism with overlap representation. 
The proton wave functions are derived from a meson-baryon fluctuation model that incorporates $|q\bar{q}B\rangle$ Fock states, providing a natural framework for investigating sea quark contributions. 
Within this approach, we compute the complete set of twist-3 GPDs and present the numerical results, including both chiral-odd ($H_2$, $E_2$, $\widetilde{E}_2$, $\widetilde{H}_2^\prime$) and chiral-even ($\widetilde{E}_{2T}$, $H_{2T}^\prime$, $E_{2T}^\prime$, $\widetilde{H}_{2T}^\prime$) distributions for $\bar{u}$ and $\bar{d}$ quarks at zero skewness. 
By taking the forward limit, we also calculate the corresponding twist-3 parton distribution functions  $e(x)$ of $\bar{u}$ and $\bar{d}$ quarks. The kinetic orbital angular momenta of sea quarks deduced from the twist-3 GPDs are studied and compared to those  from the twist-2 GPDs.
\end{abstract}

	\maketitle
	
	\section{Introduction}\label{Sec1}

Understanding the internal structure of hadrons composed of constituent quarks, gluon, and sea quarks remains a fundamental challenge in quantum chromodynamics (QCD) and hadronic physics. 
Generalized parton distributions (GPDs)~\cite{Muller:1994ses,Ji:1996nm,Radyushkin:1997ki,Diehl:2015uka} have emerged as crucial observables for probing the three-dimensional structure of nucleons, complementing the information provided by transverse momentum dependent parton distributions (TMDs). These distributions generalize conventional parton distribution functions (PDFs) to the off-forward scattering regime, naturally appearing in exclusive processes such as deeply virtual Compton scattering (DVCS), $\gamma^* h(p) \rightarrow\gamma h(p^\prime)$, and hard exclusive meson production, $\gamma^* h_1(p) \rightarrow M h_2(p^\prime)$, where the target nucleon receives recoil momentum $\Delta$. 
Significant experimental progress has been made through measurements by the H1~\cite{H1:1999pji,H1:2001nez,H1:2005gdw}, ZEUS~\cite{ZEUS:1998xpo,ZEUS:2003pwh}, HERMES~\cite{HERMES:2001bob,HERMES:2011bou,HERMES:2012gbh}, COMPASS~\cite{dHose:2004usi}, and JLab~\cite{CLAS:2001wjj} collaborations.

GPDs encode richer structural information than one-dimensional PDFs through their dependence on three variables: the longitudinal momentum fraction $x$, the momentum transfer squared $t$, and the skewness parameter $\xi$. In the forward limit ($\xi \rightarrow 0$, $t \rightarrow 0$), GPDs reduce to standard PDFs, while their integration over $x$ yields nucleon form factors. 
Notably, GPDs provide access to the quark orbital angular momentum (OAM) through the Ji sum rule~\cite{Ji:1996nm} and enable tomographic imaging of nucleon structure via impact-parameter dependent distributions obtained through Fourier transformation~\cite{Burkardt:2000za,Burkardt:2002hr,Diehl:2002he}.

The twist decomposition of GPDs, analogous to that of PDFs and TMDs, organizes contributions in inverse powers of the hard scale $Q$. 
While leading-twist GPDs dominate high-energy processes~\cite{Boffi:2007yc,Kumericki:2009uq,Burkardt:2002ks,Bhattacharya:2018zxi,Bhattacharya:2019cme}, higher-twist contributions become significant at moderate $Q^2$ values accessible in current experiments. 
These subleading terms play essential roles in preserving fundamental symmetries: leading-twist DVCS amplitudes alone violate electromagnetic Ward identities and Lorentz invariance, requiring higher-twist corrections for consistency~\cite{Anikin:2000em,Penttinen:2000dg,Belitsky:2000vx,Kivel:2000cn,Radyushkin:2000ap}. Moreover, precise extraction of twist-2 GPDs from experimental data necessitates accounting for power corrections.

In addition to correcting the leading-twist amplitudes and enabling more reliable extraction of twist-2 GPDs, there are further motivations for measuring higher-twist GPDs, particularly twist-3 ones.  
First, eight independent twist-3 GPDs contribute to nucleon DVCS at this level~\cite{Anikin:2000em,Penttinen:2000dg,Belitsky:2000vx,Kivel:2000cn,Kivel:2000fg}, with measurable effects at relatively low  $ Q^2$ values. 
Second, twist-3 distributions provide unique access to quark OAM in longitudinally polarized nucleons~\cite{Ji:1996ek,Penttinen:2000dg}, enabling decomposition of spin and OAM contributions that remain entangled in twist-2 observables~\cite{Ji:2012sj,Ji:2012ba,Hatta:2011ku,Hatta:2012cs}. 
Third, these distributions offer insights into transverse color forces~\cite{Burkardt:2008ps,Aslan:2019jis}.
In addition, certain spin-orbit correlations of the nucleon and the twist-3 DVCS amplitude can be expressed in terms of twist-3 GPDs~\cite{Lorce:2014mxa, Bhoonah:2017olu, Guo:2022cgq}. Finally, higher-twist GPDs can be related  to generalized TMDs (GTMDs)~\cite{Meissner:2008ay,Meissner:2009ww,Lorce:2013pza,Rajan:2017cpx}, which are referred to as ``mother distributions" and contain the most comprehensive structural information about hadrons.

In recent years, the kinematical twist-3 effects in the DVCS process have been studied extensively. Moreover, the twist-3 GPDs have also been studies through various approaches including quark target~\cite{Mukherjee:2002pq,Mukherjee:2002xi,Aslan:2018tff} and diquark models~\cite{Aslan:2018tff}, lattice QCD~\cite{Bhattacharya:2023nmv}, spectator models~\cite{Tan:2024doz}, light-front formalisms~\cite{Jain:2024lsj,Zhang:2023xfe}. 
However, most calculations focus on valence quark contributions, leaving sea quark twist-3 GPDs largely unexplored.

In this paper, we apply the light-cone quark model (LCQM) to calculate the twist-3 GPDs of $\bar{u}$ and $\bar{d}$ quarks at zero skewness using the overlap representation. As proposed in Refs.~\cite{Brodsky:1996hc,Pasquini:2006dv}, the sea quark degree freedom can be generated by the meson-baryon fluctuation model, where the proton can fluctuate into a composite state with a meson $M$ and a baryon $B$, and consequently the meson $M$ contains the $q\bar{q}$ Fock states. The light-cone wave functions (LCWFs) of the proton thus can be derived in terms of the $|q\bar{q}B\rangle$ Fock states, as calculated in Ref.~\cite{Luan:2022fjc}. The expressions of the twist-3 GPDs of $\bar{u}$ and $\bar{d}$ quarks, including the twist-3 chiral-odd GPDs $H_{2}^{\bar{q}/P}$, $E_{2}^{\bar{q}/P}$, $\widetilde{E}_{2}^{\bar{q}/P}$, $\widetilde{H}_{2}^{\prime \bar{q}/P}$ and twist-3 chiral-even GPDs $\widetilde{E}_{2T}^{ \bar{q}/P}$, $H_{2T}^{\prime \bar{q}/P}$, $E_{2T}^{\prime \bar{q}/P}$, $\widetilde{H}_{2T}^{\prime \bar{q}/P}$ at zero skewness, can be obtained within the overlap representation in terms of these LCWFs. The numerical results for these twist-3 GPDs are presented by properly choosing the values of the parameters. By taking the forward limit, the numerical results for the corresponding twist-3 PDFs $e^{\bar{u}/P}(x)$ and $e^{\bar{d}/P}(x)$ are also presented. 
Furthermore, the kinetic OAM of sea quarks defined by the twist-2 and twist-3 GPDs are compared.

The remained part of the paper is organized as follows. In Sec.~\ref{Sec2}, we introduce the definition of the twist-3 GPDs. We derive the overlap representation for the twist-3 GPDs in Sec.~\ref{Sec3}, where we also obtain the analytic expressions of these GPDs of the sea quarks using the LCWFs. In Sec.~\ref{Sec4}, we present the numerical results of twist-3 GPDs and PDFs. We summarize the paper in Sec.~\ref{Sec5}.

\section{Twist-3 generalized parton distributions}\label{Sec2}
    
The quark GPDs are defined through the off-forward matrix elements of the light-cone quark-quark correlation function:

	\begin{widetext}
    \begin{align}
    	F_{\Lambda^{\prime} \Lambda }^{[\Gamma]}(x, \xi, t) = \left.\frac{1}{2} \int \frac{d z^{-}}{2 \pi} e^{i k \cdot z}\left\langle p^{\prime}, \Lambda^{\prime}\left|\bar{\psi}\left(-\frac{1}{2} z\right) \Gamma \mathcal{W}\left(-\frac{1}{2} z, \left.\frac{1}{2} z \right\rvert\, n\right) \psi\left(\frac{1}{2} z\right)\right| p, \Lambda\right\rangle\right|_{z^{+} = \vec{z}_{T} = 0}.\label{eq:FLL}
    \end{align}
Here, $\mathcal{W}(-z/2, z/2|n)$ denotes the Wilson line ensuring gauge invariance,
 $p^\mu$ ($p^{\prime\mu}$) and $\Lambda$ ($\Lambda^{\prime}$) represent the four-momentum and helicity of the initial (final) nucleon state,  
and 
$$x = k^+/P^+, \,\Delta^\mu = p^{\prime\mu} - p^\mu, \,\xi = -\Delta^+/(2P^+),$$ 
are the average light-cone momentum fraction carried by the active quark, the momentum transfer with $t = \Delta^2$, and the skewness parameter, respectively.

Following the systematic classification of Ref.~\cite{Meissner:2008ay}, the complete set of twist-3 GPDs comprises sixteen independent functions, including eight chiral-odd GPDs:
\begin{align}
    	\notag
    	F_{\Lambda^{\prime} \Lambda }^{[1]} & =  \frac{M}{2\left(P^{+}\right)^{2}} \bar{u}\left(p^{\prime}, \Lambda^{\prime}\right)\left[\gamma^{+} H_{2}(x, \xi, t)+\frac{i \sigma^{+\Delta}}{2 M} E_{2}(x, \xi, t)\right] u(p, \Lambda) \\
    	&=\frac{M}{P^{+}}\left[H_{2} \delta_{\Lambda, \Lambda^{\prime}}+\frac{\left(\Lambda \Delta_{1}+i \Delta_{2}\right)}{2 M} E_{2} \delta_{-\Lambda, \Lambda^{\prime}}\right],
    	\\\notag
    	F_{\Lambda^{\prime} \Lambda }^{[\gamma_{5}]} & =  \frac{M}{2\left(P^{+}\right)^{2}} \bar{u}\left(p^{\prime}, \Lambda^{\prime}\right)\left[\gamma^{+} \gamma_{5} \tilde{H}_{2}(x, \xi, t)+\frac{P^{+} \gamma_{5}}{M} \tilde{E}_{2}(x, \xi, t)\right] u(p, \Lambda) \\
    	&=\frac{M}{P^{+}}\left[\Lambda\tilde{H}_{2} \delta_{\Lambda, \Lambda^{\prime}}-\frac{\left( \Delta_{1}+i\Lambda \Delta_{2}\right)}{2 M} \tilde{E}_{2} \delta_{-\Lambda, \Lambda^{\prime}}\right],
        \\\notag
    	F_{\Lambda^{\prime} \Lambda }^{[i \sigma^{i j} \gamma_{5}]} & =  -\frac{i \varepsilon_{T}^{i j} M}{2\left(P^{+}\right)^{2}} \bar{u}\left(p^{\prime}, \Lambda^{\prime}\right)\left[\gamma^{+} H_{2}^{\prime}(x, \xi, t)+\frac{i \sigma^{+\Delta}}{2 M} E_{2}^{\prime}(x, \xi, t)\right] u(p, \Lambda), \\
    	&=-\frac{i \varepsilon_{T}^{i j} M}{P^{+}}\left[H_{2}^{\prime} \delta_{\Lambda, \Lambda^{\prime}}+\frac{\left(\Lambda \Delta_{1}+i \Delta_{2}\right)}{2 M} E_{2}^{\prime} \delta_{-\Lambda, \Lambda^{\prime}}\right],
    	\\\notag
    	F_{\Lambda^{\prime} \Lambda }^{[i \sigma^{+-} \gamma_{5}]} & =  \frac{M}{2\left(P^{+}\right)^{2}} \bar{u}\left(p^{\prime}, \Lambda^{\prime}\right)\left[\gamma^{+} \gamma_{5} \tilde{H}_{2}^{\prime}(x, \xi, t)+\frac{P^{+} \gamma_{5}}{M} \tilde{E}_{2}^{\prime}(x, \xi, t)\right] u(p, \Lambda)\\
    	&=\frac{M}{P^{+}}\left[\Lambda\tilde{H}_{2}^{\prime} \delta_{\Lambda, \Lambda^{\prime}}-\frac{\left( \Delta_{1}+i\Lambda \Delta_{2}\right)}{2 M} \tilde{E}_{2}^{\prime} \delta_{-\Lambda, \Lambda^{\prime}}\right],
\end{align}
and eight chiral-even GPDs
\begin{align}
    	\notag
    	F_{\Lambda^{\prime} \Lambda }^{[\gamma^{i}]} & =  \frac{M}{2\left(P^{+}\right)^{2}} \bar{u}\left(p^{\prime}, \Lambda^{\prime}\right)\left[i \sigma^{+i} H_{2 T}(x, \xi, t)+\frac{\gamma^{+} \Delta_{i}-\Delta^{+} \gamma^{i}}{2 M} E_{2 T}(x, \xi, t)\right. \\\notag
    	& \left.+\frac{P^{+} \Delta_{i}-\Delta^{+} P_{i}}{M^{2}} \tilde{H}_{2 T}(x, \xi, t)+\frac{\gamma^{+} P_{i}-P^{+} \gamma^{i}}{M} \tilde{E}_{2 T}(x, \xi, t)\right] u(p, \Lambda)\\\notag
    	&=\left[\frac{\Delta_{i}}{2 P^{+}} E_{2 T}+\frac{\Delta_{i}}{P^{+}} \tilde{H}_{2 T}+\frac{i \Lambda \epsilon^{i j}_{T} \Delta_{j}}{2 P^{+}}\left(\tilde{E}_{2 T}-\xi E_{2 T}\right)\right] \delta_{\Lambda \Lambda^{\prime}} \\
    	&+\left[\frac{-M\left(\Lambda \delta_{i 1}+i \delta_{i 2}\right)}{P^{+}} H_{2 T}-\frac{\left(\Lambda \Delta_{1}+i \Delta_{2}\right) \Delta_{i}}{2 M P^{+}} \tilde{H}_{2 T}\right] \delta_{\Lambda-\Lambda^{\prime}},\\
    	\notag
    	F_{\Lambda^{\prime} \Lambda }^{[\gamma^{i} \gamma_{5}]} & =  \frac{i \varepsilon_{T}^{i j} M}{2\left(P^{+}\right)^{2}} \bar{u}\left(p^{\prime}, \Lambda^{\prime}\right)\left[i \sigma^{+j} H_{2 T}^{\prime}(x, \xi, t)+\frac{\gamma^{+} \Delta_{j}-\Delta^{+} \gamma^{j}}{2 M} E_{2 T}^{\prime}(x, \xi, t)\right. \\\notag
    	& \left.+\frac{P^{+} \Delta_{j}-\Delta^{+} P_{j}}{M^{2}} \tilde{H}_{2 T}^{\prime}(x, \xi, t)+\frac{\gamma^{+} P_{j}-P^{+} \gamma^{j}}{M} \tilde{E}_{2 T}^{\prime}(x, \xi, t)\right] u(p, \Lambda) \\\notag
    	&=\left[\frac{i \epsilon^{i j}_{T} \Delta_{j}}{2 P^{+}} E_{2 T}^{\prime}+\frac{i \epsilon^{i j}_{T} \Delta_{j}}{P^{+}} \tilde{H}_{2 T}^{\prime}-\frac{\Lambda \Delta_{i}}{2 P^{+}}\left(\tilde{E}_{2 T}^{\prime}-\xi E_{2 T}^{\prime}\right)\right] \delta_{\Lambda \Lambda^{\prime}}\\
    	&+\left[\frac{M\left(\delta_{i 1}+i \Lambda \delta_{i 2}\right)}{P^{+}} H_{2 T}^{\prime}-\frac{i \epsilon^{i j}_{T}\left(\Lambda \Delta_{1}+i \Delta_{2}\right) \Delta_{j}}{2 M P^{+}} \tilde{H}_{2 T}^{\prime}\right] \delta_{\Lambda-\Lambda^{\prime}},
\end{align}
where $\sigma^{i j}=i\left[\gamma^{i}, \gamma^{j}\right] / 2 $ and $\epsilon_{T}^{i j}=\epsilon^{-+i j}$. 
Among them, $\tilde{H}_{2}, H_{2}^{\prime}, E_{2}^{\prime}, \tilde{E}_{2}^{\prime}, H_{2 T}, E_{2 T}, \tilde{H}_{2 T}, \tilde{E}_{2 T}^{\prime}$ are odd functions of $\xi$, while $H_{2}, E_{2}, \widetilde{E}_{2}, \tilde{H}_{2}^{\prime}, \widetilde{E}_{2T},H_{2T}^{\prime},E_{2T}^{\prime},\widetilde{H}_{2T}^{\prime}$ are even functions of $\xi$.

A different parametrization with vector and axial-vector components is introduced in Ref.~\cite{Aslan:2018zzk}, where $F^{\mu}$-type and $\tilde{F}^{\mu}$-type GPDs are parameterized as:
\begin{align}
    	F^{\mu} & = \bar{u}\left(p^{\prime}\right)\left[P^{\mu} \frac{\gamma^{+}}{P^{+}} H+P^{\mu} \frac{i \sigma^{+\nu} \Delta_{\nu}}{2 M P^{+}} E+\Delta_{T}^{\mu} \frac{1}{2 M} G_{1}+\gamma_{T}^{\mu}\left(H+E+G_{2}\right)+\Delta_{T}^{\mu} \frac{\gamma^{+}}{P^{+}} G_{3}+i \epsilon_{T}^{\mu \nu} \Delta_{\nu} \frac{\gamma^{+} \gamma_{5}}{P^{+}} G_{4}\right] u(p), \\
    	\tilde{F}^{\mu} & = \bar{u}\left(p^{\prime}\right)\left[P^{\mu} \frac{\gamma^{+} \gamma_{5}}{P^{+}} \tilde{H}+P^{\mu} \frac{\Delta^{+} \gamma_{5}}{2 M P^{+}} \tilde{E}+\Delta_{T}^{\mu} \frac{\gamma_{5}}{2 M}\left(\tilde{E}+\tilde{G}_{1}\right)+\gamma_{T}^{\mu} \gamma_{5}\left(\tilde{H}+\tilde{G}_{2}\right)+\Delta_{T}^{\mu} \frac{\gamma^{+} \gamma_{5}}{P^{+}} \tilde{G}_{3}+i \epsilon_{T}^{\mu \nu} \Delta_{\nu} \frac{\gamma^{+}}{P^{+}} \tilde{G}_{4}\right] u(p) .
\end{align}
And the GPDs from these two different parameterizations can be connected through the following relations~\cite{Aslan:2018zzk}
    \begin{align}
    	G_{1} & = 2 \tilde{H}_{2 T}, & G_{2} & = -(H+E)-\frac{1}{\xi}\left(1-\xi^{2}\right) H_{2 T}+\xi E_{2 T}-\tilde{E}_{2 T}, \\
    	G_{3} & = \frac{1}{2}\left(H_{2 T}+E_{2 T}\right), &G_{4} &= \frac{1}{2 \xi} H_{2 T}, \\
    	\tilde{G}_{1} & = -\tilde{E}+2 \tilde{H}_{2 T}^{\prime}, &\tilde{G}_{2}& = -\tilde{H}+\left(1-\xi^{2}\right) H_{2 T}^{\prime}-\xi^{2} E_{2 T}^{\prime}-\frac{\Delta_{T}^{2}}{2 M^{2}} \tilde{H}_{2 T}^{\prime}+\xi \tilde{E}_{2 T}^{\prime}, \\
    	\tilde{G}_{3} & = -\frac{\xi}{2}\left(H_{2 T}^{\prime}+E_{2 T}^{\prime}\right)-\frac{\xi (M^{2}+t/4)}{M^{2}} \tilde{H}_{2 T}^{\prime}+\frac{1}{2} \tilde{E}_{2 T}^{\prime}, &\tilde{G}_{4} &= -\frac{1}{2}\left(H_{2 T}^{\prime}+E_{2 T}^{\prime}\right)-\frac{M^{2}+t/4}{M^{2}} \tilde{H}_{2 T}^{\prime}.
    \end{align}
        
Using $+$ ($-$) to denote the positive (negative) helicity of the proton, the twist-3 chiral-odd GPDs ($i=1$, $j=2$), can be expressed as the combinations of the helicity amplitudes of the correlator (\ref{eq:FLL}):
   \begin{align} 
   	\frac{M}{P^{+}}{H}_{2}&=\frac{F^{[1]}_{++}+F^{[1]}_{--}}{2}, 	
   	&
   	\frac{i\Delta_2}{2P^{+}}{E}_{2}&=\frac{F^{[1]}_{+-}+F^{[1]}_{-+}}{2},
   	\\
   	\frac{M}{P^{+}}\widetilde{H}_{2}&=\frac{F^{[\gamma_{5}]}_{++}	-F^{[\gamma_{5}]}_{--}}{2}, 
   	&	-\frac{\Delta_1}{2P^{+}}\widetilde{E}_{2}&=\frac{F^{[\gamma_{5}]}_{+-}+F^{[\gamma_{5}]}_{-+}}{2}, 
   	\\
   	-\frac{iM}{P^{+}}{H}_{2}^{\prime}&=\frac{F^{[i \sigma^{1 2} \gamma_{5}]}_{++}	+F^{[i \sigma^{1 2} \gamma_{5}]}_{--}}{2}, 	
   	&
   	\frac{\Delta_2}{2P^{+}}{E}_{2}^{\prime}&=\frac{F^{[i \sigma^{1 2} \gamma_{5}]}_{+-}+F^{[i \sigma^{1 2} \gamma_{5}]}_{-+}}{2},
   	\\
   	\frac{M}{P^{+}}\widetilde{H}_{2}^{\prime}&=\frac{F^{[i \sigma^{+-} \gamma_{5}]}_{++}	-F^{[i \sigma^{1 2} \gamma_{5}]}_{--}}{2}, 	
   	&
   	-\frac{\Delta_1}{2P^{+}}\widetilde{E}_{2}^{\prime}&=\frac{F^{[i \sigma^{+-} \gamma_{5}]}_{+-}+F^{[i \sigma^{1 2} \gamma_{5}]}_{-+}}{2}.	  
   \end{align} 	

Similarly, the twist-3 chiral-even GPDs can be expressed as
\begin{align} 	
\frac{\boldsymbol{\Delta}_T^2}{2P^{+}}(E_{2T}+2\widetilde{H}_{2T})&=\frac{\Delta_1(F^{[\gamma^{1}]}_{++}+F^{[\gamma^{1}]}_{--})}{2}+\frac{\Delta_2(F^{[\gamma^{2}]}_{++}+F^{[\gamma^{2}]}_{--})}{2},	
   	\\
   	\frac{i\boldsymbol{\Delta}_T^2}{2P^{+}}(\widetilde{E}_{2T}-\xi E_{2T})&=\frac{\Delta_2(F^{[\gamma^{1}}_{++}-F^{[\gamma^{1}]}_{--})}{2}-\frac{\Delta_1(F^{[\gamma^{2}]}_{++}-F^{[\gamma^{2}]}_{--})}{2},  
   	\\
   	\frac{2M}{P^{+}}({H}_{2T}+\frac{\boldsymbol{\Delta}_T^2}{4M^2}\widetilde{H}_{2T})&=\frac{F^{[\gamma^{1}]}_{+-}-F^{[\gamma^{1}]}_{-+}}{2}+\frac{i(F^{[\gamma^{2}]}_{+-}+F^{[\gamma^{2}]}_{-+})}{2},
   	\\
   	\frac{i\Delta_1\Delta_2}{MP^{+}}\widetilde{H}_{2T}&=\frac{-(F^{[\gamma^{1}]}_{+-}+F^{[\gamma^{1}]}_{-+})}{2}+\frac{i(F^{[\gamma^{2}]}_{+-}-F^{[\gamma^{2}]}_{-+})}{2},  	
   	\\
   	\frac{i\boldsymbol{\Delta}_T^2}{2P^{+}}(E_{2T}^{\prime}+2\widetilde{H}_{2T}^{\prime})&=\frac{\Delta_2(F^{[\gamma^{1}\gamma_{5}]}_{++}+F^{[\gamma^{1}\gamma_{5}]}_{--})}{2}-\frac{\Delta_1(F^{[\gamma^{2}\gamma_{5}]}_{++}+F^{[\gamma^{2}\gamma_{5}]}_{--})}{2},  
   	\\
   	\frac{-\boldsymbol{\Delta}_T^2}{2P^{+}}(\widetilde{E}_{2T}^{\prime}-\xi{E}_{2T}^{\prime})&=\frac{\Delta_1(F^{[\gamma^{1}\gamma_{5}]}_{++}-F^{[\gamma^{1}\gamma_{5}]}_{--})}{2}+\frac{\Delta_2(F^{[\gamma^{2}\gamma_{5}]}_{++}-F^{[\gamma^{2}\gamma_{5}]}_{--})}{2},
   	\\
   	\frac{2M}{P^{+}}({H}_{2T}^{\prime}+\frac{\boldsymbol{\Delta}_T^2}{4M^2}\widetilde{H}_{2T}^{\prime})&=\frac{F^{[\gamma^{1}\gamma_{5}]}_{+-}+F^{[\gamma^{1}\gamma_{5}]}_{-+}}{2}+\frac{i(F^{[\gamma^{2}\gamma_{5}]}_{+-}-F^{[\gamma^{2}\gamma_{5}]}_{-+})}{2},
   	\\
   	\frac{i\Delta_1\Delta_2}{MP^{+}}\widetilde{H}_{2T}^{\prime}&=\frac{F^{[\gamma^{1}\gamma_{5}]}_{+-}-F^{[\gamma^{1}\gamma_{5}]}_{-+}}{2}-\frac{i(F^{[\gamma^{2}\gamma_{5}]}_{+-}+F^{[\gamma^{2}\gamma_{5}]}_{-+})}{2}.
   \end{align}

\section{Twist-3 GPDs of sea quarks in the overlap representation}\label{Sec3}

In this section, we present the calculation of twist-3 GPDs for $\bar{u}$ and $\bar{d}$ sea quarks in the proton at zero skewness ($\xi=0$) using the LCQM with overlap representation. 
The light-cone formalism has proven particularly powerful for studying hadron structure, having been successfully applied to calculations of nucleon and meson parton distribution functions~\cite{Lepage:1980fj}, form factors~\cite{Brodsky:2000ii,Xiao:2003wf}, anomalous magnetic moments~\cite{Brodsky:2000ii}, TMDs~\cite{Bacchetta:2008af,Lu:2006kt,Luan:2022fjc}, GPDs~\cite{Muller:2014tqa,Brodsky:2000xy,Burkardt:2003je,Luan:2023lmt}, as well as quark Wigner distributions~\cite{Ma:2018ysi,Kaur:2020par,Luan:2024nwc}. 
We now extend this framework to investigate the twist-3 GPDs of sea quarks.

In the light-cone approach, the hadronic state is expanded in terms of light-cone wave functions (LCWFs) on a Fock-state basis. To incorporate sea quark degrees of freedom, we employ the baryon-meson fluctuation model~\cite{Brodsky:1996hc,Pasquini:2006dv}, where the proton can fluctuate into a meson-baryon two-body system:
\begin{align}\label{fock state}
    	|p\rangle\to| M B\rangle\to|q\bar{q}B\rangle,
\end{align}
where the meson is composed of the $q\bar{q}$ pair.

In our analysis, we consider the specific fluctuations $|p\rangle \rightarrow |\pi^+ n\rangle$ and $|p\rangle \rightarrow |\pi^- \Delta^{++}\rangle$. The corresponding LCWFs, derived in Ref.~\cite{Luan:2022fjc}, factorize into baryonic and mesonic components:
\begin{align}\label{LCWFs} \psi^{\lambda_N}_{{\lambda_B}{\lambda_q}{\lambda_{\bar{q}}}}(x,y,\boldsymbol{k}_T,\boldsymbol{r}_T)		=&\psi^{\lambda_N}_{\lambda_B}(y,\boldsymbol{r}_T)\psi_{{\lambda_q}{\lambda_{\bar{q}}}}
    	(x,y,\boldsymbol{k}_T,\boldsymbol{r}_T),
\end{align}
where $\psi^{\lambda_N}_{\lambda_B}(y,r_T)$ can be viewed as the wave function of the nucleon in terms $\pi B$ components, and $\psi_{{\lambda_q}{\lambda_{\bar{q}}}}
(x,y,\boldsymbol{k}_T,\boldsymbol{r}_T)$ is the pion wave function in terms of $q \bar{q}$ components. The indices $\lambda_N$, $\lambda_B$, $\lambda_q$, $\lambda_{\bar{q}}$ denote the helicities of the proton, the baryon, the quark and the sea quark, respectively. 
$x$ and $y$ represent their light-cone momentum fractions, $\boldsymbol{k}_T$ and $\boldsymbol{r}_T$ denote the transverse momenta of the antiquark and the meson. 
    
The baryonic wave functions for different helicity combinations are given by:
\begin{align}\label{former}
    	\notag\psi^+_+(y,\boldsymbol{r}_T)&=\frac{M_B-(1-y)M}{\sqrt{1-y}}\phi_1, \\
    	\notag\psi^+_-(y,\boldsymbol{r}_T)&=\frac{r_1+ir_2}{\sqrt{1-y}}\phi_1, \\
    	\notag\psi^-_+(y,\boldsymbol{r}_T)&=\frac{r_1-ir_2}{\sqrt{1-y}}\phi_1 , \\
    	\psi^-_-(y,\boldsymbol{r}_T)&=\frac{(1-y)M-M_B}{\sqrt{1-y}}\phi_1,
\end{align}	
where $M$ and $M_B$ are proton and baryon masses respectively, and $\phi_1$ is the momentum-space wave function:
\begin{equation}
    \phi_1(y,\bm{r}_T) = -\frac{g(r^2)\sqrt{y(1-y)}}{\bm{r}_T^2 + L_1^2(m_\pi^2)},
\end{equation}
with $m_\pi$ the pion mass, $g(r^2)$ is the form factor for the coupling of the nucleon-pion meson-baryon vertex, and
\begin{equation}
    L_1^2(m_\pi^2) = yM_B^2 + (1-y)m_\pi^2 - y(1-y)M^2.
\end{equation}

The mesonic wave functions for the $q\bar{q}$ Fock state are:
\begin{align}\label{later}		\notag\psi{_+}{_+}(x,y,\boldsymbol{k}_T,\boldsymbol{r}_T)&=\frac{my}{\sqrt{x(y-x)}}\phi_2,\\ \notag\psi{_+}{_-}(x,y,\boldsymbol{k}_T,\boldsymbol{r}_T)&=\frac{y(k_1-ik_2)-x(r_1-ir_2)}{\sqrt{x(y-x)}}\phi_2, \\ \notag\psi{_-}{_+}(x,y,\boldsymbol{k}_T,\boldsymbol{r}_T)&=\frac{y(k_1+ik_2)-x(r_1+ir_2)}{\sqrt{x(y-x)}}\phi_2, \\		\psi{_-}{_-}(x,y,\boldsymbol{k}_T,\boldsymbol{r}_T)&=\frac{-my}{\sqrt{x(y-x)}}\phi_2.
    \end{align} 
Here, $m$ is the mass of quark and the sea quark. 
Again, $\phi_2(x,y,\boldsymbol{k}_T,\boldsymbol{r}_T)$ is the momentum space wave function of the $|q\bar{q}\rangle$ Fock state 
\begin{align}	     	
    \phi_2(x,y,\boldsymbol{k}_T,\boldsymbol{r}_T)=
    	-\frac{g(k^2)\sqrt{\frac{x}{y}(1-\frac{x}{y})}}{(\boldsymbol{k}_T
    		-\frac{x}{y}\boldsymbol{r}_T)^2+L_2^2(m^2)},
\end{align}
$g(k^2)$ is the form factor for the coupling of the pion meson-quark-sea quark vertex, and
\begin{align}		
    	L_2^2(m^2)=\frac{x}{y}m^2+\left(1-\frac{x}{y}\right)m^2
    	-\frac{x}{y}\left(1-\frac{x}{y}\right){m_\pi}^2.
\end{align}
    
For the form factors $g(r^2)$ and $g(k^2)$, we adopt the dipolar form
\begin{align}
    	g(r^2)&=-g_1(1-y)\frac{\boldsymbol{r}_T^2+L_1^2(m_\pi^2)}
    	{[\boldsymbol{r}_T^2+L_1^2(\Lambda^2_\pi)]^2},\label{eq15}	\\		g(k^2)&=-g_2(1-\frac{x}{y})\frac{(\boldsymbol{k}_T-\frac{x}{y}\boldsymbol{r}_T)^2+L_2^2(m^2)}{[(\boldsymbol{k}_T-\frac{x}{y}\boldsymbol{r}_T)^2+L_2^2(\Lambda^2_{\bar{q}})]^2}.\label{eq16}	
\end{align}

When considering the quark as the active parton (with the antiquark as spectator), the quark carries momentum $-\bm{k}_T$ and longitudinal momentum fraction $-x$, while the spectator antiquark has momentum $(y+x, \bm{r}_T + \bm{k}_T)$. The corresponding LCWFs maintain consistent relations under this transformation. 
\begin{align}\label{relation1}	\notag\psi{_+}{_+}(-x,y,-\boldsymbol{k}_T,\boldsymbol{r}_T)&=
    	\psi{_+}{_+}(x,y,\boldsymbol{k}_T,\boldsymbol{r}_T),\\	\notag\psi{_+}{_-}(-x,y,-\boldsymbol{k}_T,\boldsymbol{r}_T)&=-
    	\psi{_+}{_-}(x,y,\boldsymbol{k}_T,\boldsymbol{r}_T), \\      	\notag\psi{_-}{_+}(-x,y,-\boldsymbol{k}_T,\boldsymbol{r}_T)&=-
    	\psi{_-}{_+}(x,y,\boldsymbol{k}_T,\boldsymbol{r}_T), \\   	\psi{_-}{_-}(-x,y,-\boldsymbol{k}_T,\boldsymbol{r}_T)&=\psi{_-}{_-}(x,y,\boldsymbol{k}_T,\boldsymbol{r}_T).
\end{align} 
The antiquark distributions are defined by using this conjugate correlation function, which can be related to quark-quark correlation function by the following expression~\cite{Kumano:2021fem}
    \begin{align}\label{relation2}
    	\notag&F^{C[\Gamma]}_{\Lambda^{\prime}\Lambda}(x, \xi, t; m_{q},m_{\bar{q}})
    	\\& =\left\{\begin{array}{ll}
    		-F^{[\Gamma]}_{\Lambda^{\prime}\Lambda}(-x, \xi, t; m_{\bar{q}},m_{q})  \text { for } \Gamma =\gamma^{j}, \gamma^{-}, i \sigma^{i j} \gamma_{5}, i \sigma^{j-} \gamma_{5} \\
    		+F^{[\Gamma]}_{\Lambda^{\prime}\Lambda}(-x, \xi, t; m_{\bar{q}},m_{q})  \text { for } \Gamma = \mathbf{1}, \gamma^{j} \gamma_{5}
    	\end{array}\right.
\end{align}

In the overlap representation, the quark-quark correlation function at $\xi=0$ can be expressed as~\cite{Sharma:2023ibp}
\begin{align}\label{correlation}
    	\notag F^{[\Gamma]}_{\Lambda^{\prime}\Lambda}\left(x,0, t\right)  &=\sum_{\lambda_{i}} \int [d x][d^{2} k_{T}] \psi^{\Lambda^{\prime}*}_{\lambda_{1}^{\prime} \lambda_{i}^{\prime}}\left(x_{i}, \boldsymbol{k}_{T}^{i\prime}\right) \psi^{\Lambda}_{\lambda_{1} \lambda_{i}}\left(x_{i}, \boldsymbol{k}_{T}^{i},\right)[\delta_{\lambda_i^\prime,\lambda_i}(i=2...n)]
    	\\
    	&\times\frac{u^{\dagger}\left(x_{1} P^{+}, \boldsymbol{k}_{T}^{1}-\frac{\boldsymbol{\Delta}_{T}}{2},\lambda_{1}^{\prime}\right) \gamma^{0} \Gamma u\left(x_{1} P^{+}, \boldsymbol{k}_{T}^
    		{1}+\frac{\boldsymbol{\Delta}_{T}}{2},\lambda_{1}\right)}{2 x P^{+}} ,
    \end{align}
where $\lambda_1(\lambda_1^\prime)$ represents the helicity of the initial(final) struck quark, and $\lambda_i(\lambda_i^\prime) (i=2...n)$ denotes the helicity of the initial(final) spectators. 
The spinor product corresponds to the higher-twist Dirac matrices and encodes struck quark helicity combinations. And
\begin{align}
{[d x][d^{2} k_{T}] } = \prod \frac{d x_{i} \boldsymbol{k}_{T}^{i}}{16 \pi^{3}} 16 \pi^{3} \delta\left(1-\sum x_{i}\right) \delta^{2}_{T}\left(\sum \boldsymbol{k}_{T}^{i}\right) \delta\left(x-x_{1}\right).
\end{align}
\end{widetext}

Using the LCWFs in Eqs.~(\ref{former})(\ref{later}) as well as the overlap representation in Eqs.~(\ref{odd})-(\ref{even}), we obtain the analytic results for the twist-3 chiral-odd GPDs of sea quarks
\begin{widetext}
    \begin{align}
    	\notag H_2^{\overline{q}/P}(x,0,t)&=\frac{m}{xM}\int_{x}^{1}\frac{dy}{y}\int d^2\boldsymbol{k}_T\int{d^2\boldsymbol{r}_T} \frac{g_1^2g_2^2y(1-y)^2(1-\frac{x}{y})^2 \left[(M_B-(1-y)M)^2+\boldsymbol{r}_T^2-\frac{1}{4}(1-y)\boldsymbol{\Delta}_T^2\right]}{2(2\pi)^6 D_1(y,\boldsymbol{r}_T,\boldsymbol{\Delta}_T) D_2(\frac{x}{y},\boldsymbol{k}_T-\frac{x}{y}\boldsymbol{r}_T,\boldsymbol{\Delta}_T)}
    	\\&\times  \left[m^2+(\boldsymbol{k}_T-\frac{x}{y}  \boldsymbol{r}_T)^2-\frac{1}{4}(1-\frac{x}{y})^2\boldsymbol{\Delta}_T^2+(1-\frac{x}{y})\frac{\boldsymbol{\Delta}_T^2}{2}\right],
    	\\\notag       	E_2^{\overline{q}/P}(x,0,t)&=-\frac{m}{x}\int_{x}^{1}\frac{dy}{y}\int d^2\boldsymbol{k}_T\int{d^2\boldsymbol{r}_T} \frac{g_1^2g_2^2y(1-y)^3(1-\frac{x}{y})^2 (M_B-(1-y)M)}{(2\pi)^6 D_1(y,\boldsymbol{r}_T,\boldsymbol{\Delta}_T) D_2(\frac{x}{y},\boldsymbol{k}_T-\frac{x}{y}\boldsymbol{r}_T,\boldsymbol{\Delta}_T)}
    	\\&\times	 \left[m^2+(\boldsymbol{k}_T-\frac{x}{y}  \boldsymbol{r}_T)^2-\frac{1}{4}(1-\frac{x}{y})^2\boldsymbol{\Delta}_T^2+\frac{1}{2}(1-\frac{x}{y})\boldsymbol{\Delta}_T^2\right],
    	\\      
    	\widetilde{E}_{2}^{\overline{q}/P}(x,0,t)&=0,
    	\\      	\widetilde{H}_{2}^{\prime\overline{q}/P}(x,0,t)&=\frac{m}{Mx}\int_{x}^{1}\frac{dy}{y}\int d^2\boldsymbol{k}_T\int{d^2\boldsymbol{r}_T} \frac{g_1^2g_2^2x(1-y)^3(1-\frac{x}{y})^3 (\boldsymbol{r}_T\times\boldsymbol{\Delta}_T)^2}{2(2\pi)^6 D_1(y,\boldsymbol{r}_T,\boldsymbol{\Delta}_T) D_2(\frac{x}{y},\boldsymbol{k}_T-\frac{x}{y}\boldsymbol{r}_T,\boldsymbol{\Delta}_T)},
    \end{align}
and the twist-3 chiral-even GPDs of sea quarks
	\begin{align}
       \notag       	\widetilde{E}_{2T}^{\overline{q}/P}(x,0,t)&=-\frac{1}{x}\int_{x}^{1}\frac{dy}{y}\int d^2\boldsymbol{k}_T\int{d^2\boldsymbol{r}_T} \frac{g_1^2g_2^2y(1-y)^3(1-\frac{x}{y})^2 (\boldsymbol{r}_T\times\boldsymbol{\Delta}_T)}{2(2\pi)^6 D_1(y,\boldsymbol{r}_T,\boldsymbol{\Delta}_T) D_2(\frac{x}{y},\boldsymbol{k}_T-\frac{x}{y}\boldsymbol{r}_T,\boldsymbol{\Delta}_T)}
    	\\&\times	 \left[2\left(m^2+(\boldsymbol{k}_T-\frac{x}{y}  \boldsymbol{r}_T)^2-\frac{1}{4}(1-\frac{x}{y})^2\boldsymbol{\Delta}_T^2\right)\frac{\boldsymbol{k}_T\times\boldsymbol{\Delta}_T}{\boldsymbol{\Delta}_T^{2}}+(1-\frac{x}{y})(\boldsymbol{k}_T-\frac{x}{y}  \boldsymbol{r}_T)\times\boldsymbol{\Delta}_{T}\right],
    	\\\notag       	
    	E_{2T}^{\prime \overline{q}/P}(x,0,t)&=\frac{1}{x}\int_{x}^{1}\frac{dy}{y}\int d^2\boldsymbol{k}_T\int{d^2\boldsymbol{r}_T} \frac{g_1^2g_2^2y(1-y)^2(1-\frac{x}{y})^2 }{2(2\pi)^6 D_1(y,\boldsymbol{r}_T,\boldsymbol{\Delta}_T) D_2(\frac{x}{y},\boldsymbol{k}_T-\frac{x}{y}\boldsymbol{r}_T,\boldsymbol{\Delta}_T)}
    	\\\notag&\times
    	\left[(M_B-(1-y)M)^2+\boldsymbol{r}_T^2-\frac{1}{4}(1-y)\boldsymbol{\Delta}_T^2-2(1-y)M(M_B-(1-y)M)\right]
    	\\&\times	 \left[\left(m^2+(\boldsymbol{k}_T-\frac{x}{y}  \boldsymbol{r}_T)^2-\frac{1}{4}(1-\frac{x}{y})^2\boldsymbol{\Delta}_T^2-2(1-\frac{x}{y})m^2\right)-2(1-\frac{x}{y})((\boldsymbol{k}_T-\frac{x}{y}  \boldsymbol{r}_T)\times\boldsymbol{\Delta}_{T})\frac{\boldsymbol{k}_T\times\boldsymbol{\Delta}_T}{\boldsymbol{\Delta}_T^{2}}\right],
    	\\\notag       	
    	\widetilde{H}_{2T}^{\prime \overline{q}/P}(x,0,t)&=\frac{1}{x}\int_{x}^{1}\frac{dy}{y}\int d^2\boldsymbol{k}_T\int{d^2\boldsymbol{r}_T} \frac{g_1^2g_2^2y(1-y)^3(1-\frac{x}{y})^2M\left(M_{B}-(1-y)M\right)}{2(2\pi)^6 D_1(y,\boldsymbol{r}_T,\boldsymbol{\Delta}_T) D_2(\frac{x}{y},\boldsymbol{k}_T-\frac{x}{y}\boldsymbol{r}_T,\boldsymbol{\Delta}_T)}
    	\\&\times	 \left[\left(m^2+(\boldsymbol{k}_T-\frac{x}{y}  \boldsymbol{r}_T)^2-\frac{1}{4}(1-\frac{x}{y})^2\boldsymbol{\Delta}_T^2-2(1-\frac{x}{y})m^2\right)-2(1-\frac{x}{y})((\boldsymbol{k}_T-\frac{x}{y}  \boldsymbol{r}_T)\times\boldsymbol{\Delta}_{T})\frac{\boldsymbol{k}_T\times\boldsymbol{\Delta}_T}{\boldsymbol{\Delta}_T^{2}}\right],
    	\\      	
    	H_{2T}^{\prime \overline{q}/P}(x,0,t)&=0,
    \end{align}
    where
    \begin{align}      	 D_1(y,\boldsymbol{r}_T,\boldsymbol{\Delta}_T)&=\left[(\boldsymbol{r}_T-\frac{1}{2}(1-y)
    	\boldsymbol{\Delta}_T)^2+L_1^2\right]^2 \left[(\boldsymbol{r}_T+\frac{1}{2}(1-y)\boldsymbol{\Delta}_T)^2+L_1^2\right]^2,\\
    	D_2(\frac{x}{y},\boldsymbol{k}_T-\frac{x}{y}\boldsymbol{r}_T,\boldsymbol{\Delta}_T)
    	&=\left[[(\boldsymbol{k}_T-\frac{x}{y}\boldsymbol{r}_T)-\frac{1}{2}(1-\frac{x}{y})
    	\boldsymbol{\Delta}_T]^2+L_2^2\right]^2
    	\left[[(\boldsymbol{k}_T-\frac{x}{y}\boldsymbol{r}_T)
    	+\frac{1}{2}(1-\frac{x}{y})\boldsymbol{\Delta}_T]^2+L_2^2\right]^2.
    \end{align}
   \end{widetext}

\section{Numerical results and discussion}\label{Sec4}
    
In this section, we present the numerical results for twist-3 GPDs of the $\bar{u}$ and $\bar{d}$ quarks. For the parameters $g_1$, $g_2$, $\Lambda_{\bar{q}}$, $\Lambda_\pi$ in our model, we adopt the values from Ref.~\cite{Luan:2022fjc}, where $g_2$ and $\Lambda_\pi$ are fixed by adopting the GRV leading-order (LO) parametrization~\cite{Gluck:1991ey} to perform the fit for $f_1^{\bar{u}/\pi^-}$ (or $f_1^{\bar{d}/\pi^+}(x)$). The MSTW2008 LO parametrization~\cite{Martin:2009iq} is adopted for $f_1^{\bar{u}/P}$ and $f_1^{\bar{d}/P}$ to obtain the values of the parameters $g_1$ and $\Lambda_{\bar{q}}$. 
The best values from the fits of the parameters are shown in Table.~\ref{tab1}.

\subsection{ Results for twist-3 GPDs of sea quarks}

    \begin{center}\label{tab1}
    	\setlength{\tabcolsep}{5mm}
    	\renewcommand\arraystretch{1.5}
    	\begin{tabular}{ c | c | c }
    		\hline
    		Parameters & $\bar{u}$ & $\bar{d}$ \\
    		\hline
    		\hline
    		$g_1$ & 9.33 & 5.79 \\
    		\hline
    		$g_2$ & 4.46 & 4.46 \\
    		\hline
    		$\Lambda_\pi(GeV)$ &  0.223 & 0.223 \\
    		\hline
    		$\Lambda_{\bar{q}}(GeV)$ &  0.510 &  0.510 \\
    		\hline
    	\end{tabular}
    	\captionof{table}{Values of the parameters obtained in Ref.~\cite{Luan:2022fjc}.} \label{tab1}
    \end{center}
    
    \begin{figure*}[htbp]
    	\centering
    	\subfigure{\begin{minipage}[b]{0.45\linewidth}
    			\centering
    			\includegraphics[width=\linewidth]{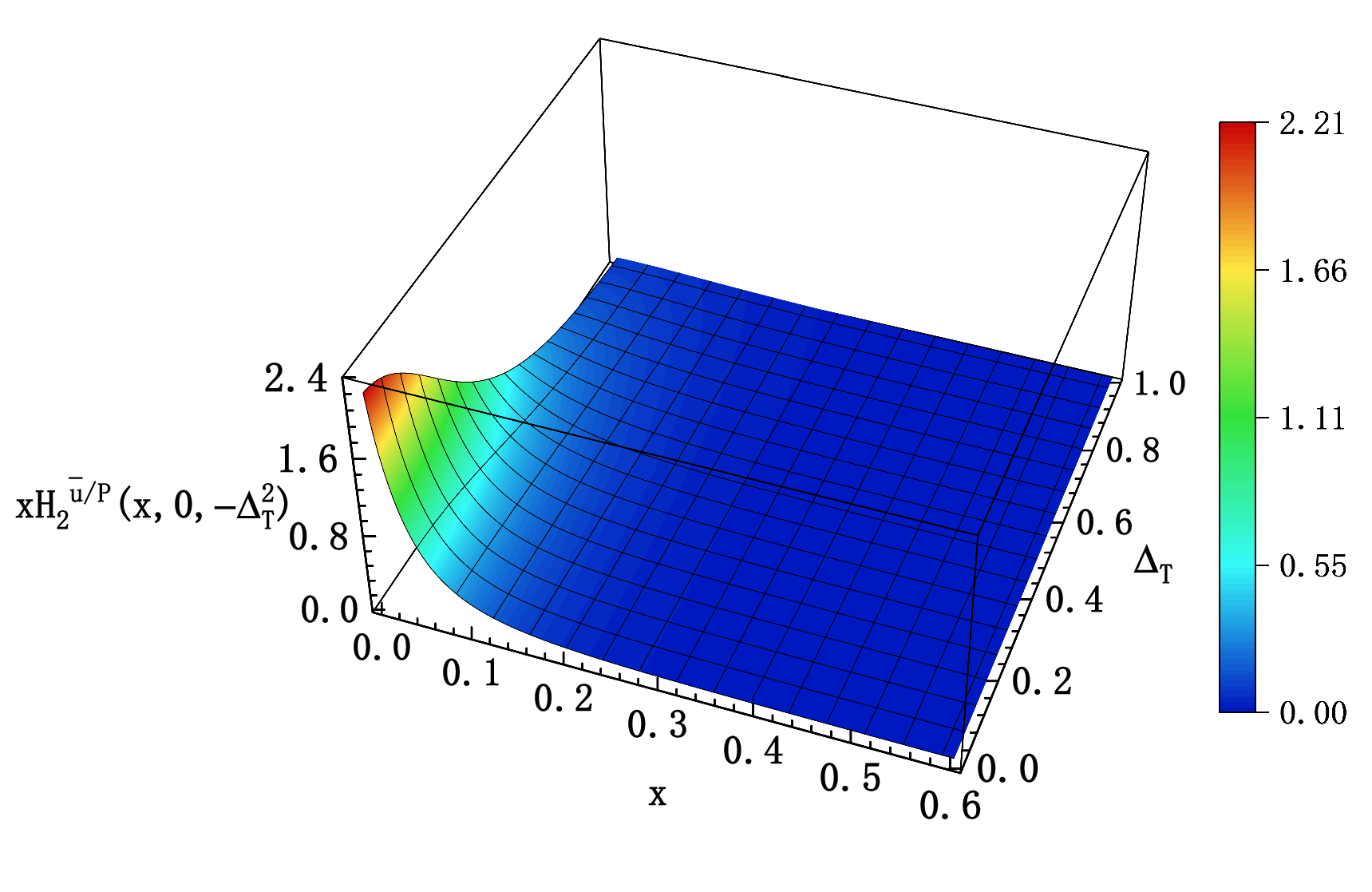}
    	\end{minipage}}
    	\subfigure{\begin{minipage}[b]{0.45\linewidth}
    			\centering
    			\includegraphics[width=\linewidth]{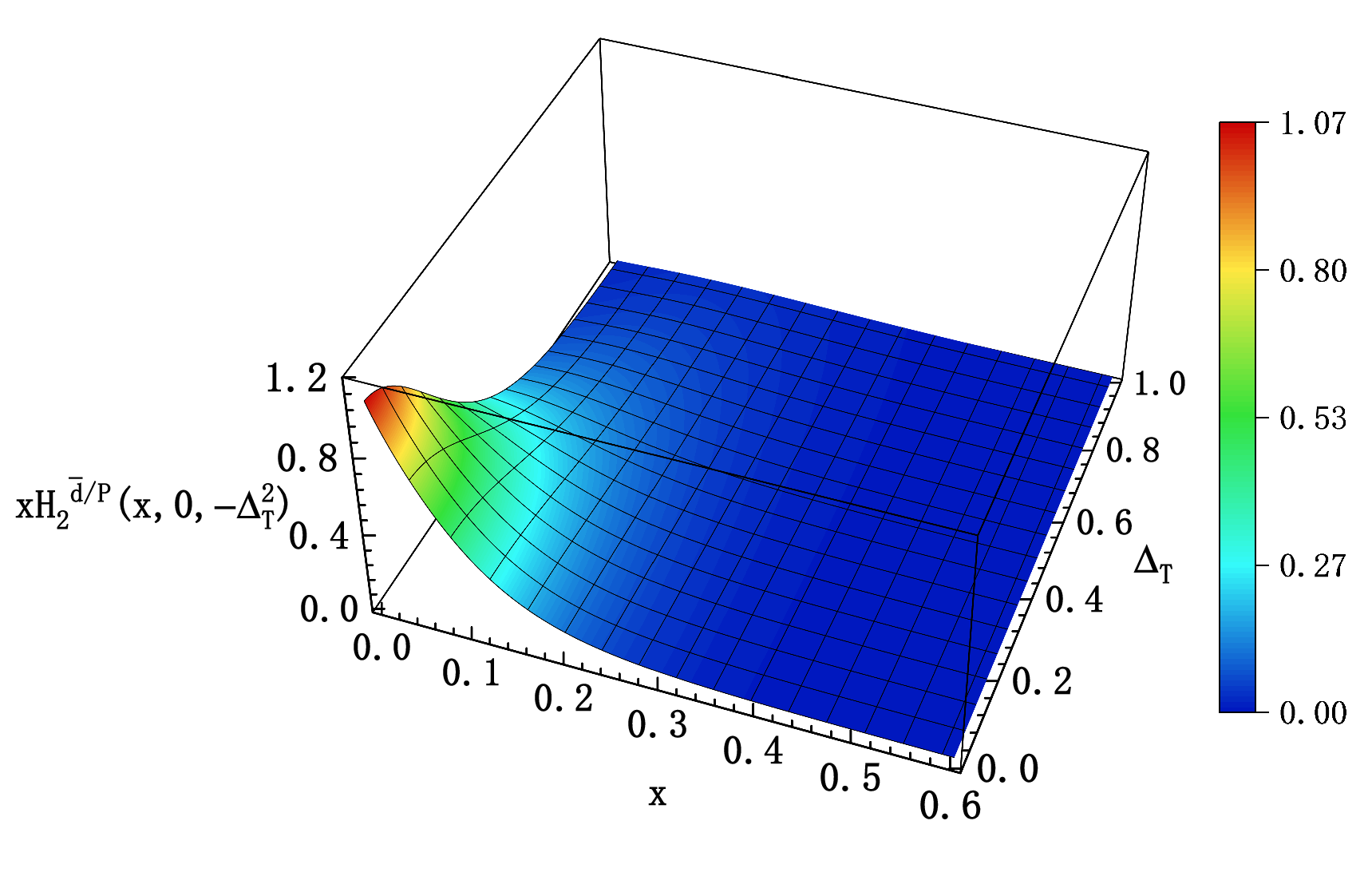}    	\end{minipage}}
    	\subfigure{\begin{minipage}[b]{0.45\linewidth}
    			\centering
    			\includegraphics[width=\linewidth]{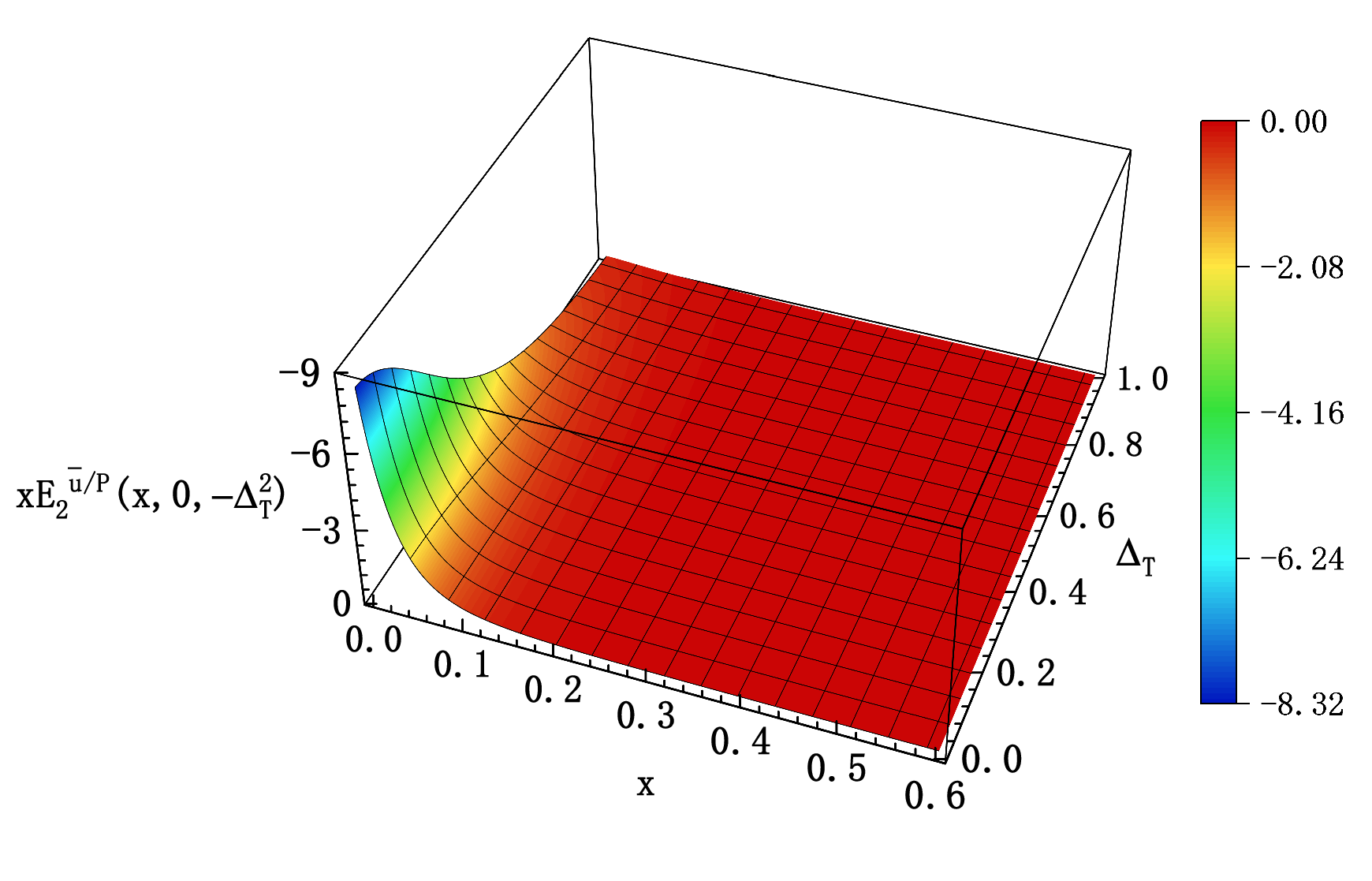}
    	\end{minipage}}
    	\subfigure{\begin{minipage}[b]{0.45\linewidth}
    			\centering
    			\includegraphics[width=\linewidth]{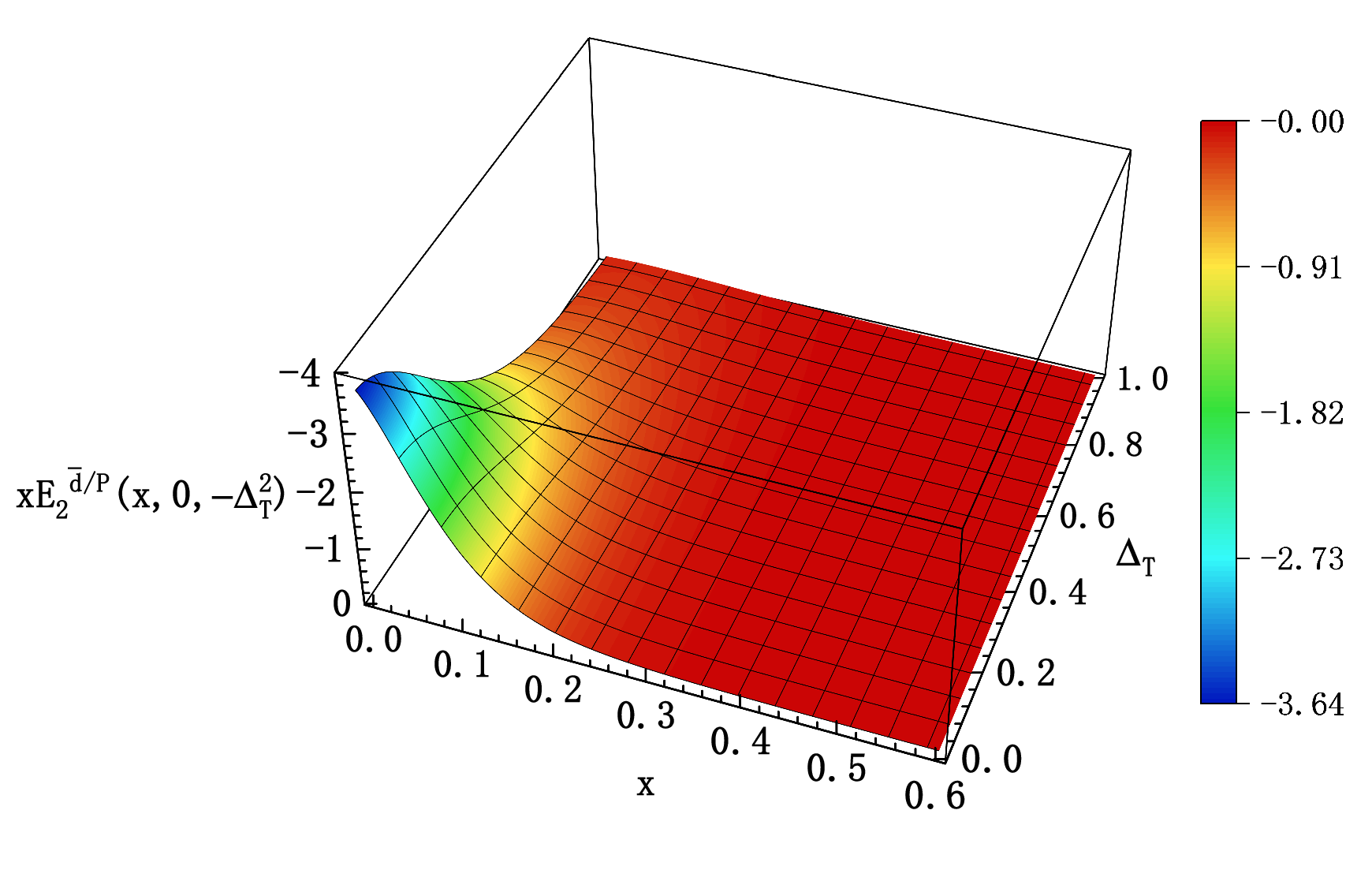}    	\end{minipage}}	
    	\caption{The twist-3 chiral-even GPDs (multiplied with $x$) $H_2^{\bar{u}/P}$, $H_2^{\bar{d}/P}$, $E_2^{\bar{u}/P}$ and $E_2^{\bar{d}/P}$, in the light-cone quark model as functions of $x$ and $\Delta_T$.} \label{H2E2}      
    \end{figure*}
  
    \begin{figure*}[htbp]
    	\centering
    	\subfigure{\begin{minipage}[b]{0.45\linewidth}
    			\centering
    			\includegraphics[width=\linewidth]{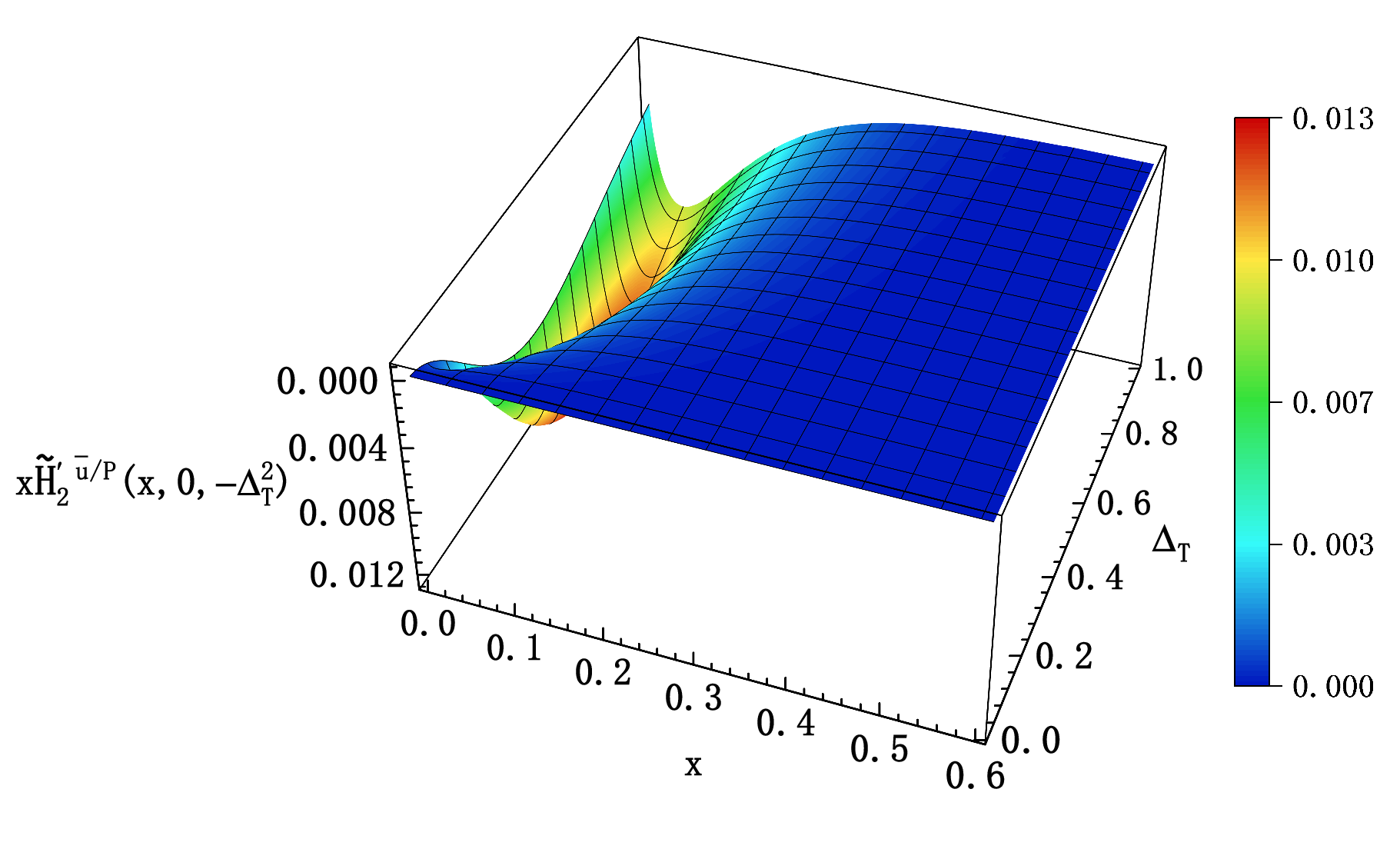}
    	\end{minipage}}
    	\subfigure{\begin{minipage}[b]{0.45\linewidth}
    			\centering
    			\includegraphics[width=\linewidth]{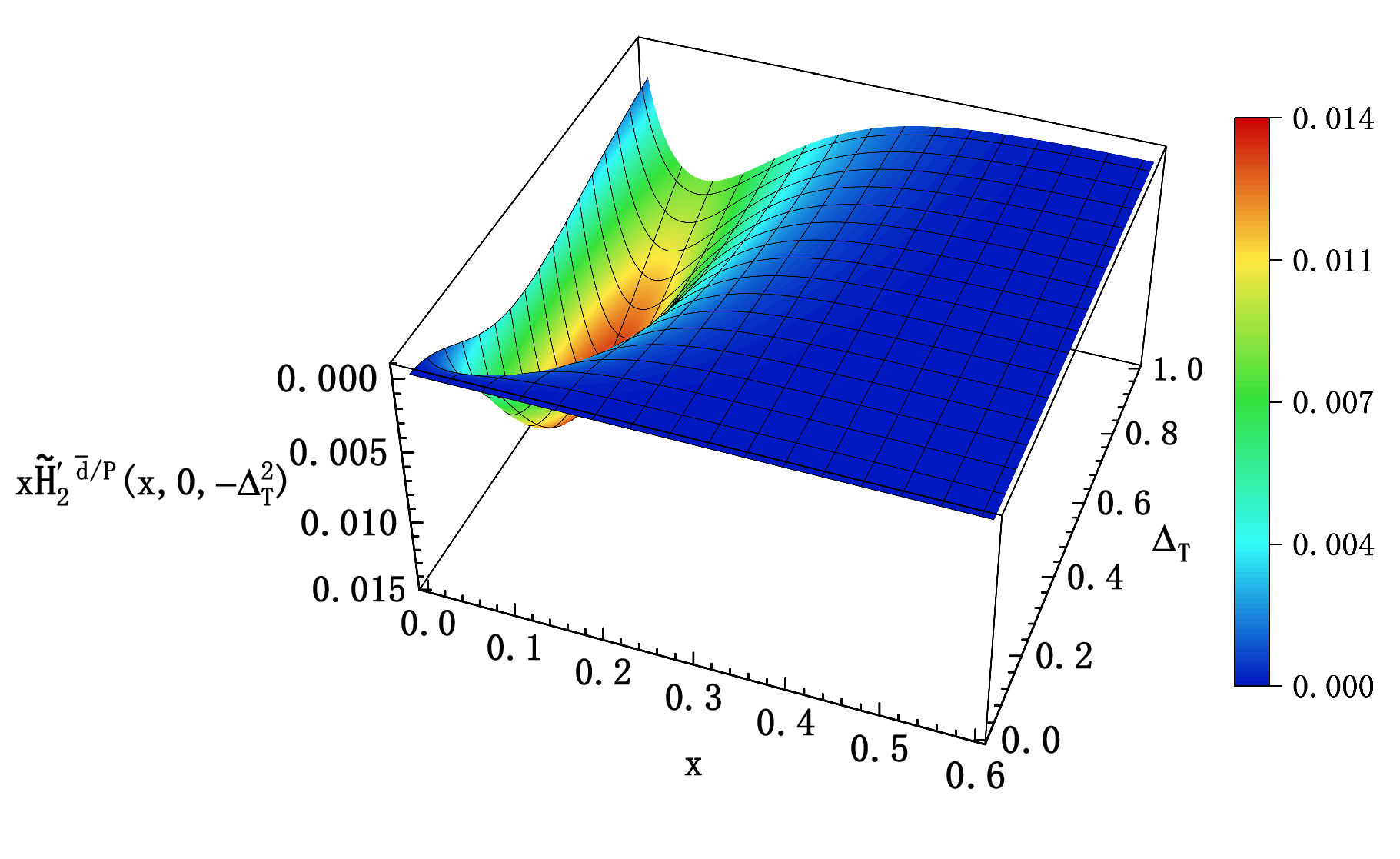}    	\end{minipage}}
    	\caption{The twist-3 chiral-even GPDs (multiplied with $x$) $\widetilde{H}_2^{\prime \bar{u}/P}$ and $\widetilde{H}_2^{\prime \bar{d}/P}$, in the light-cone quark model as functions of $x$ and $\Delta_T$.} \label{wH2p}      
    \end{figure*}

    \begin{figure*}[htbp]
    	\centering
    	\subfigure{\begin{minipage}[b]{0.45\linewidth}
    			\centering
    			\includegraphics[width=\linewidth]{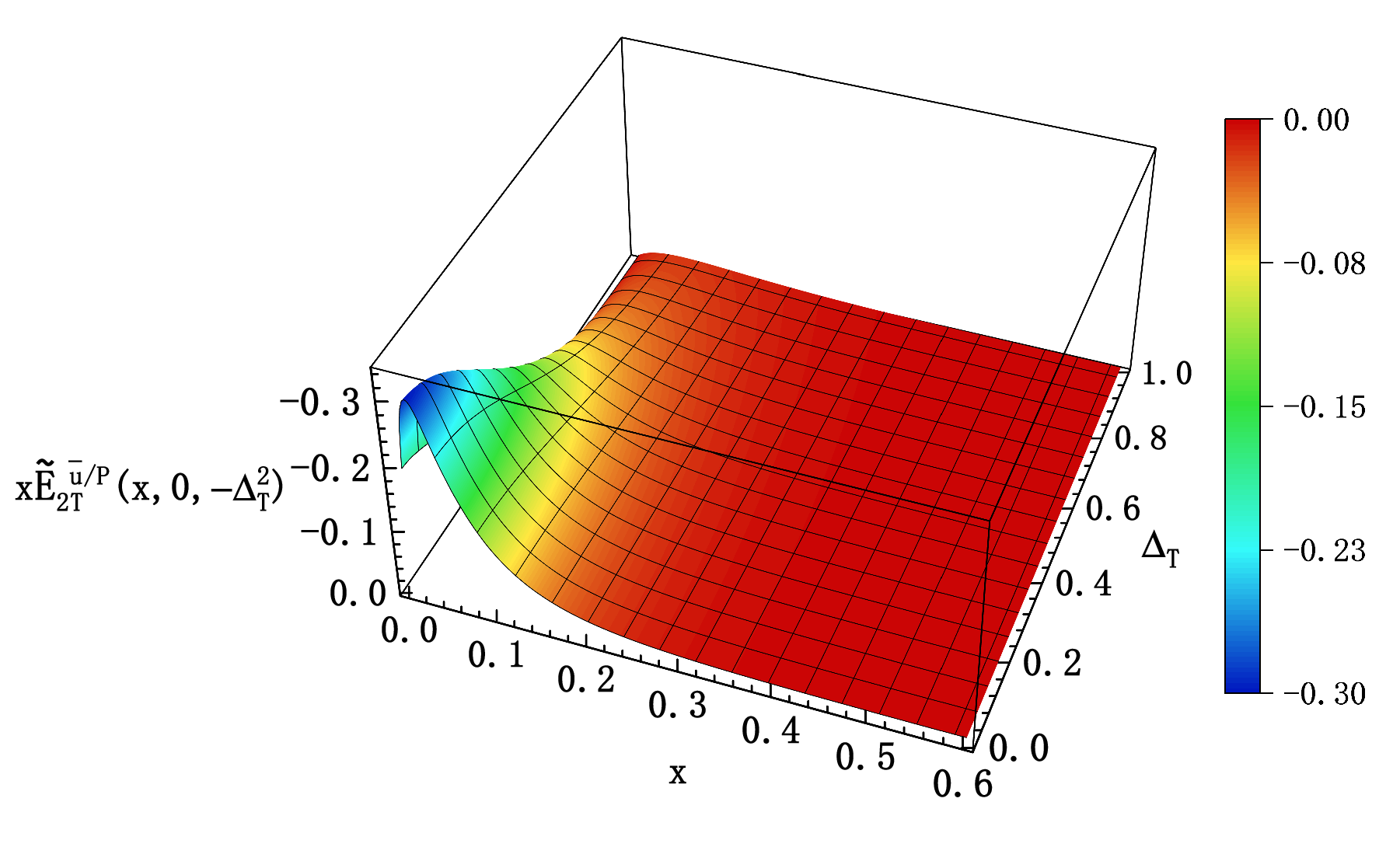}
    	\end{minipage}}
    	\subfigure{\begin{minipage}[b]{0.45\linewidth}
    			\centering
    			\includegraphics[width=\linewidth]{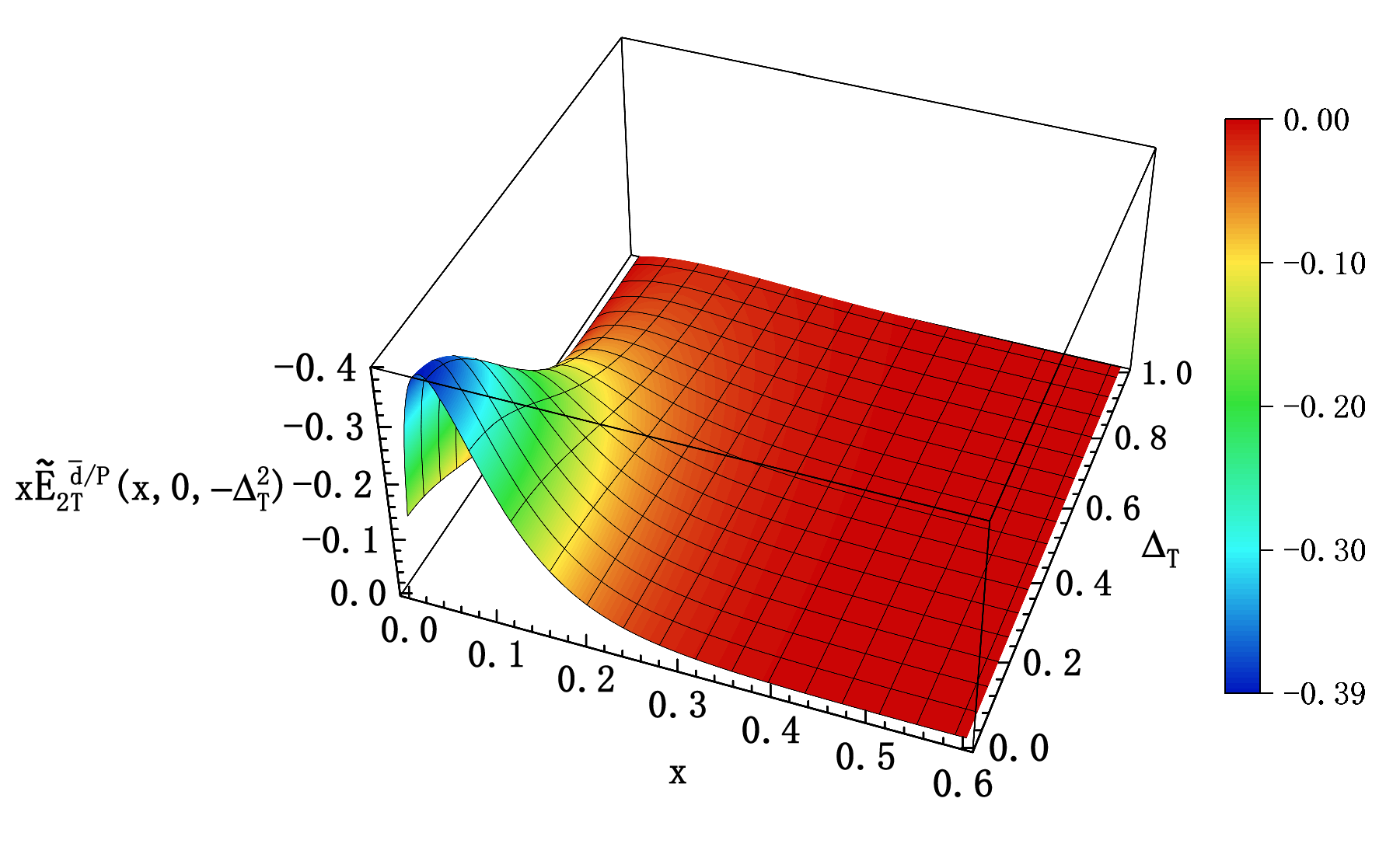}    	\end{minipage}}
    	\caption{The twist-3 chiral-odd GPDs (multiplied with $x$) $\widetilde{E}^{\bar{u}/P}_{2T}$ and $\widetilde{E}^{\bar{d}/P}_{2T}$, in the light-cone quark model as functions of $x$ and $\Delta_T$.} \label{wE2T}      
    \end{figure*}

     \begin{figure*}[htbp]
    	\centering
    	\subfigure{\begin{minipage}[b]{0.45\linewidth}
    			\centering
    			\includegraphics[width=\linewidth]{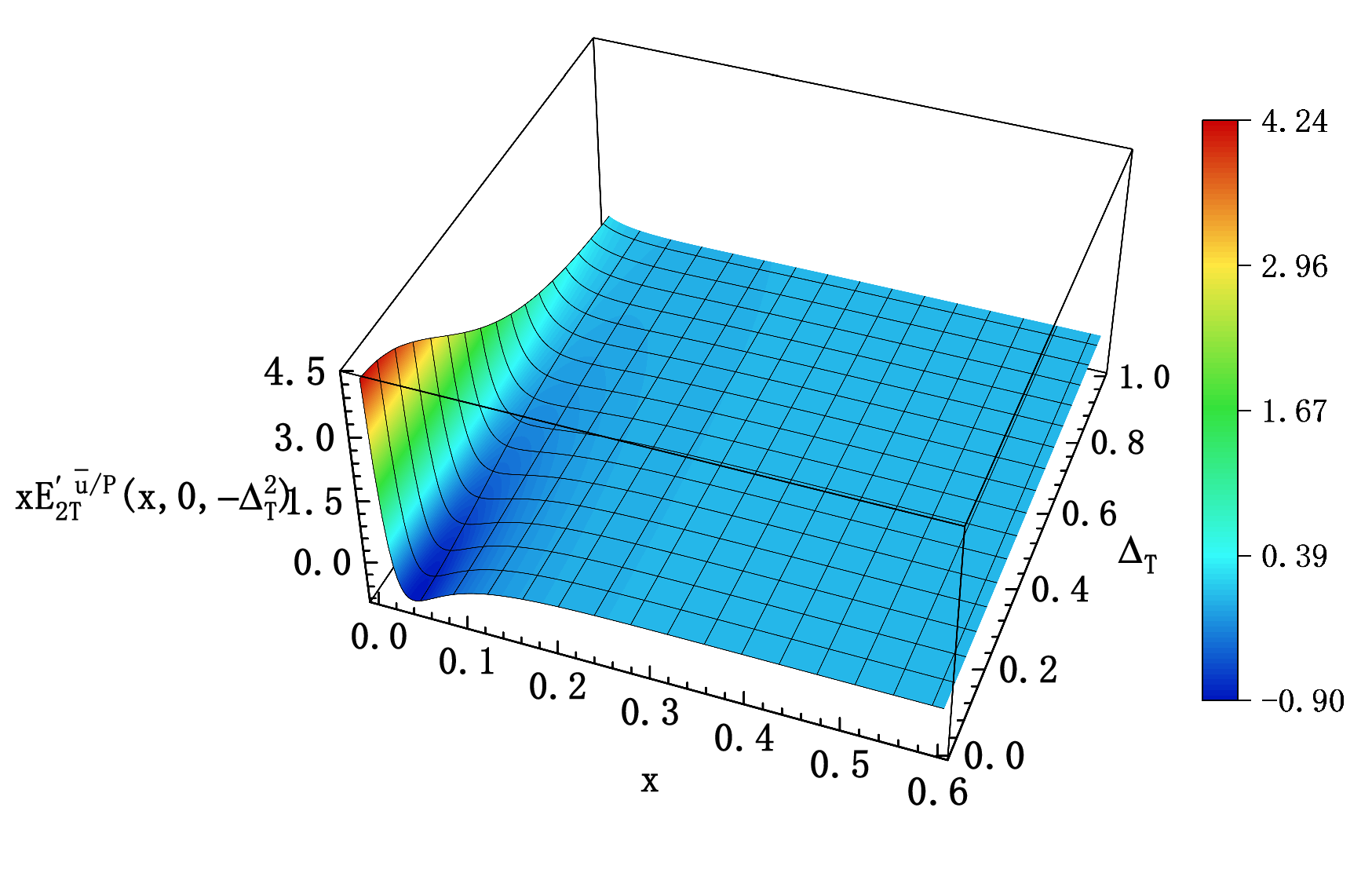}
    	\end{minipage}}
    	\subfigure{\begin{minipage}[b]{0.45\linewidth}
    			\centering
    			\includegraphics[width=\linewidth]{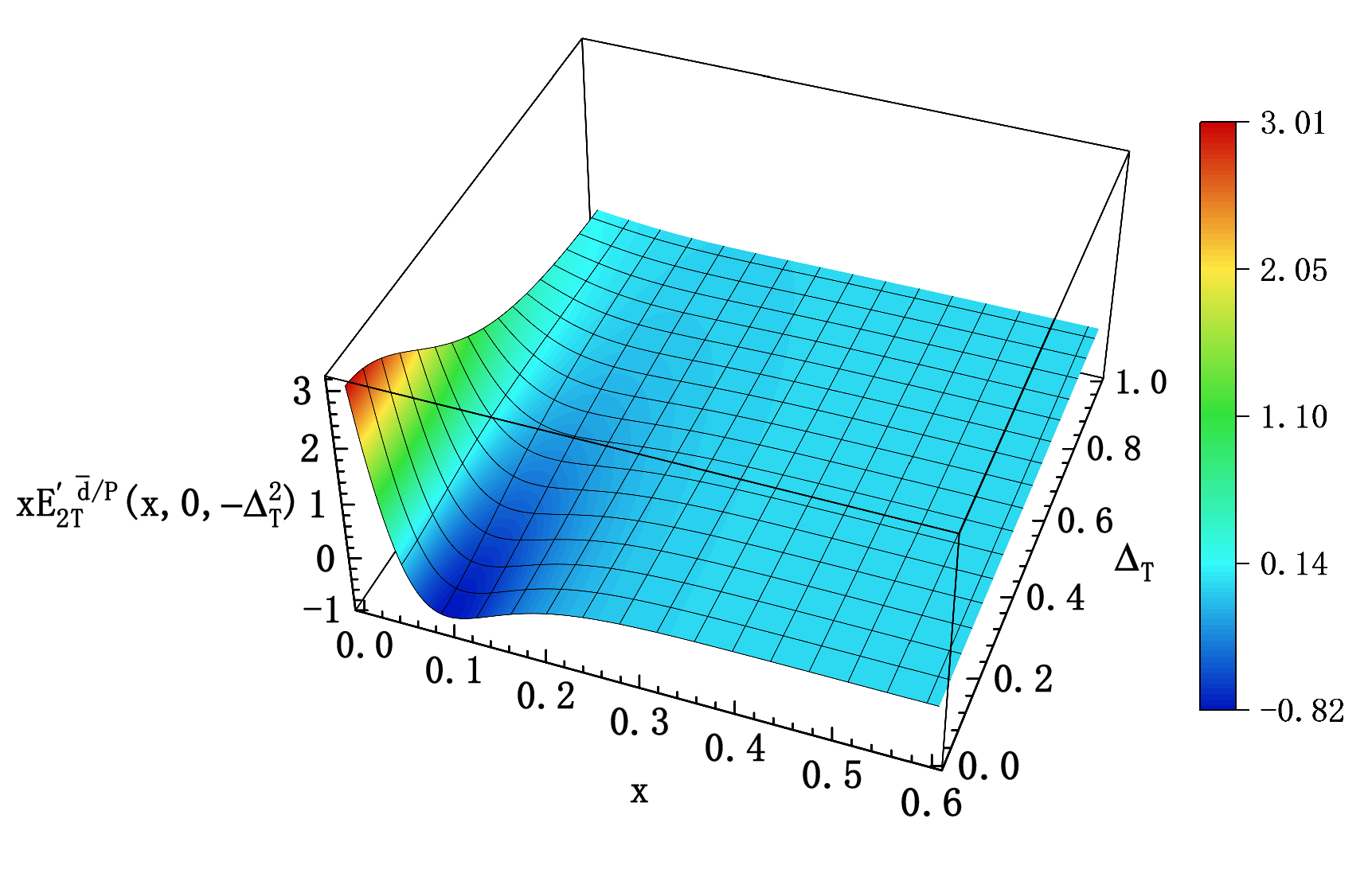}    	\end{minipage}}
    	\subfigure{\begin{minipage}[b]{0.45\linewidth}
    			\centering
    			\includegraphics[width=\linewidth]{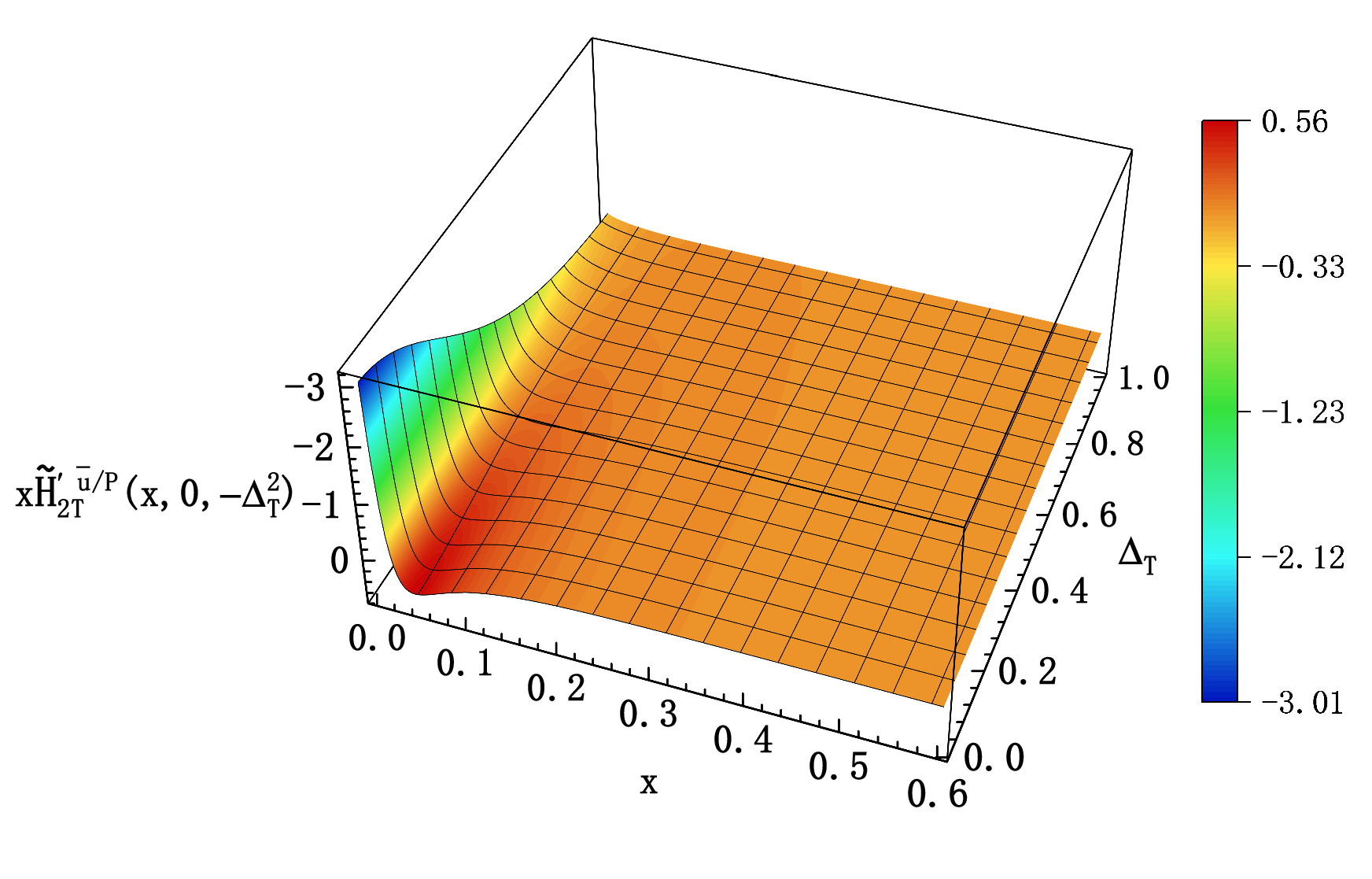}
    	\end{minipage}}
    	\subfigure{\begin{minipage}[b]{0.45\linewidth}
    			\centering
    			\includegraphics[width=\linewidth]{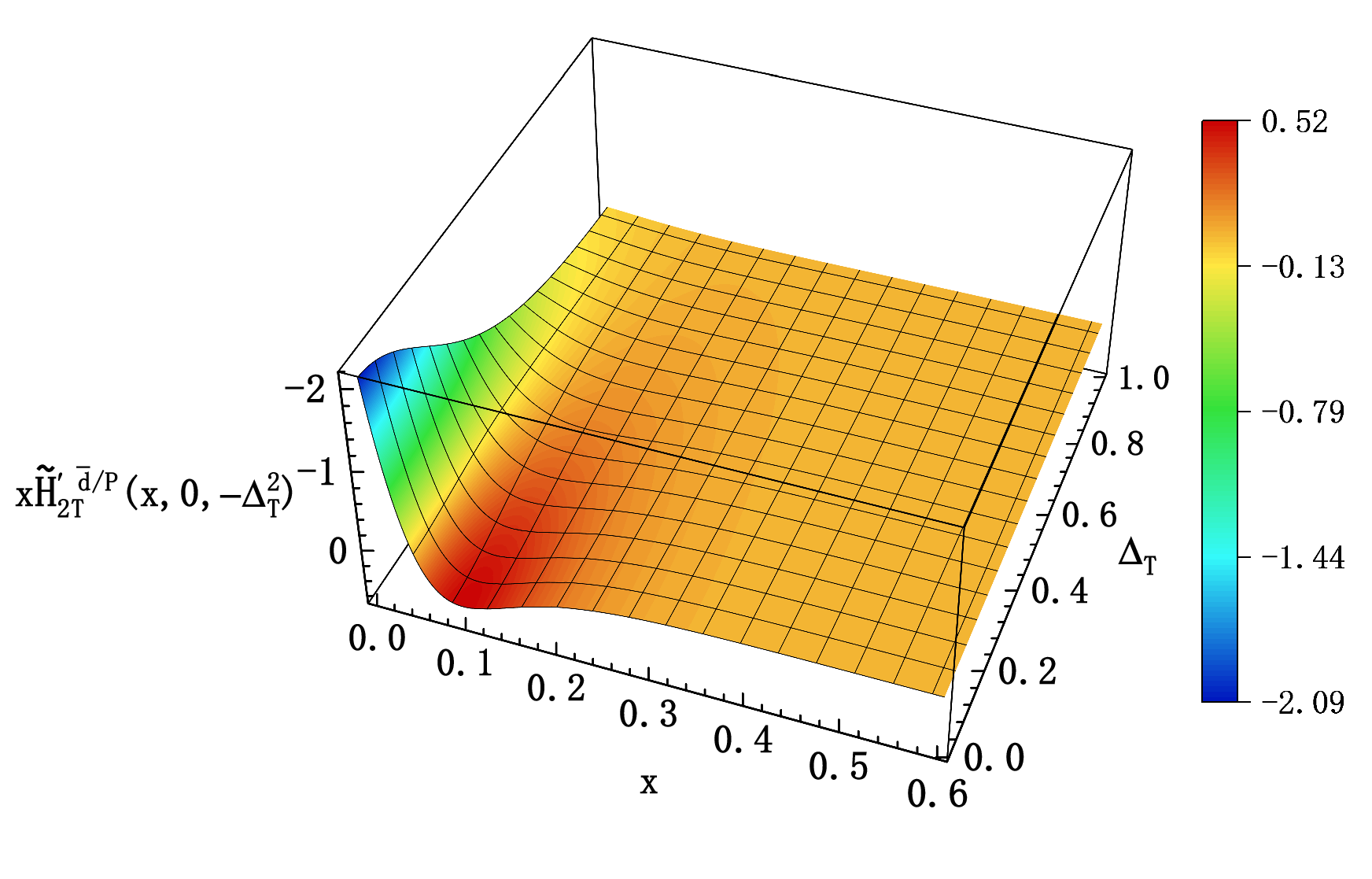}    	\end{minipage}}
    	\caption{The twist-3 chiral-odd GPDs (multiplied with $x$) $E^{\prime\bar{u}/P}_{2T}$, $E^{\prime\bar{d}/P}_{2T}$, $\widetilde{H}^{\prime\bar{u}/P}_{2T}$ and $\widetilde{H}^{\prime\bar{d}/P}_{2T}$, in the light-cone quark model as functions of $x$ and $\Delta_T$.} \label{E2TpwH2Tp}      
    \end{figure*}

     \begin{figure*}[htbp]
     	\centering
     	\subfigure{\begin{minipage}[b]{0.45\linewidth}
     			\centering
     			\includegraphics[width=\linewidth]{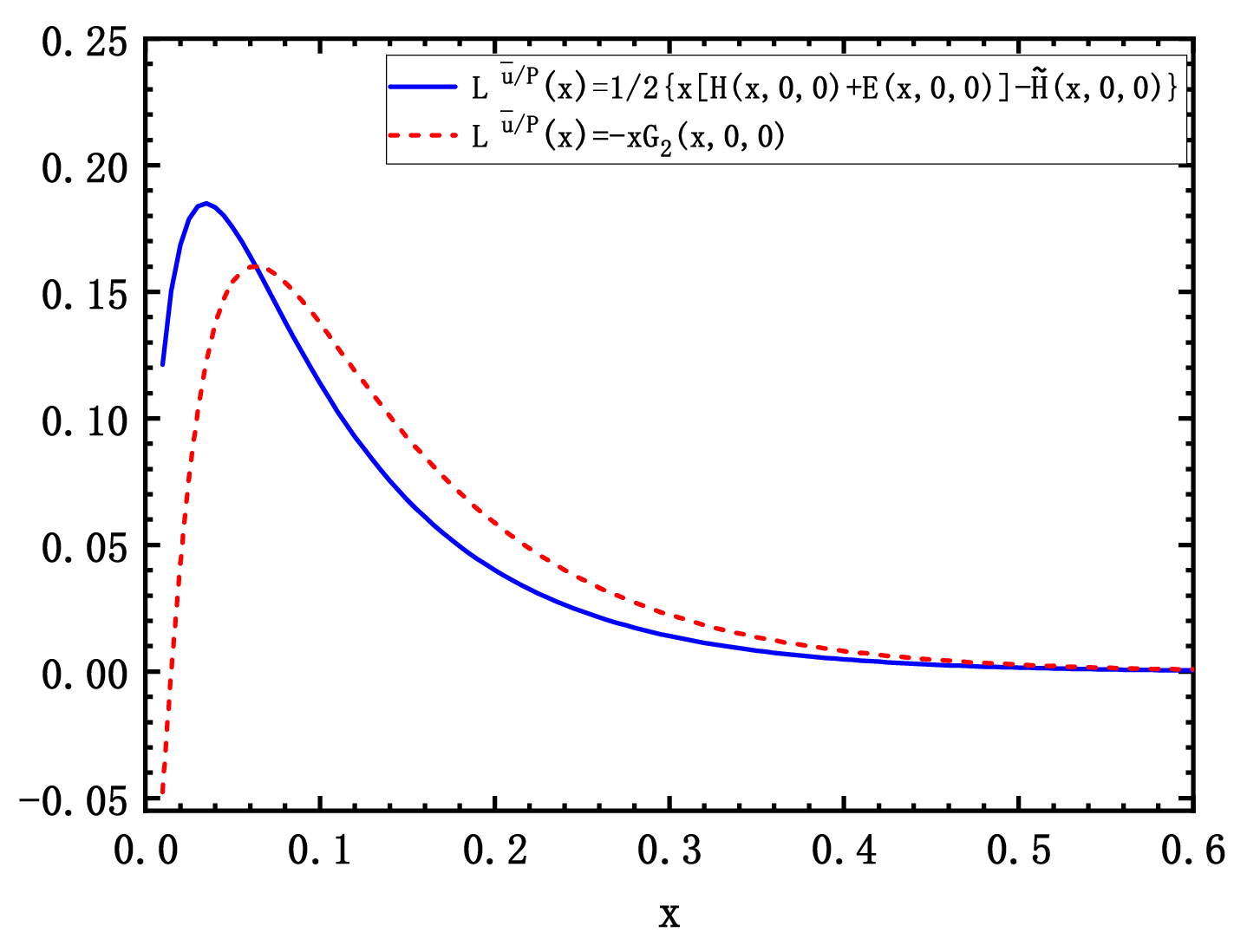}
     	\end{minipage}}
     	\subfigure{\begin{minipage}[b]{0.45\linewidth}
     			\centering
     			\includegraphics[width=\linewidth]{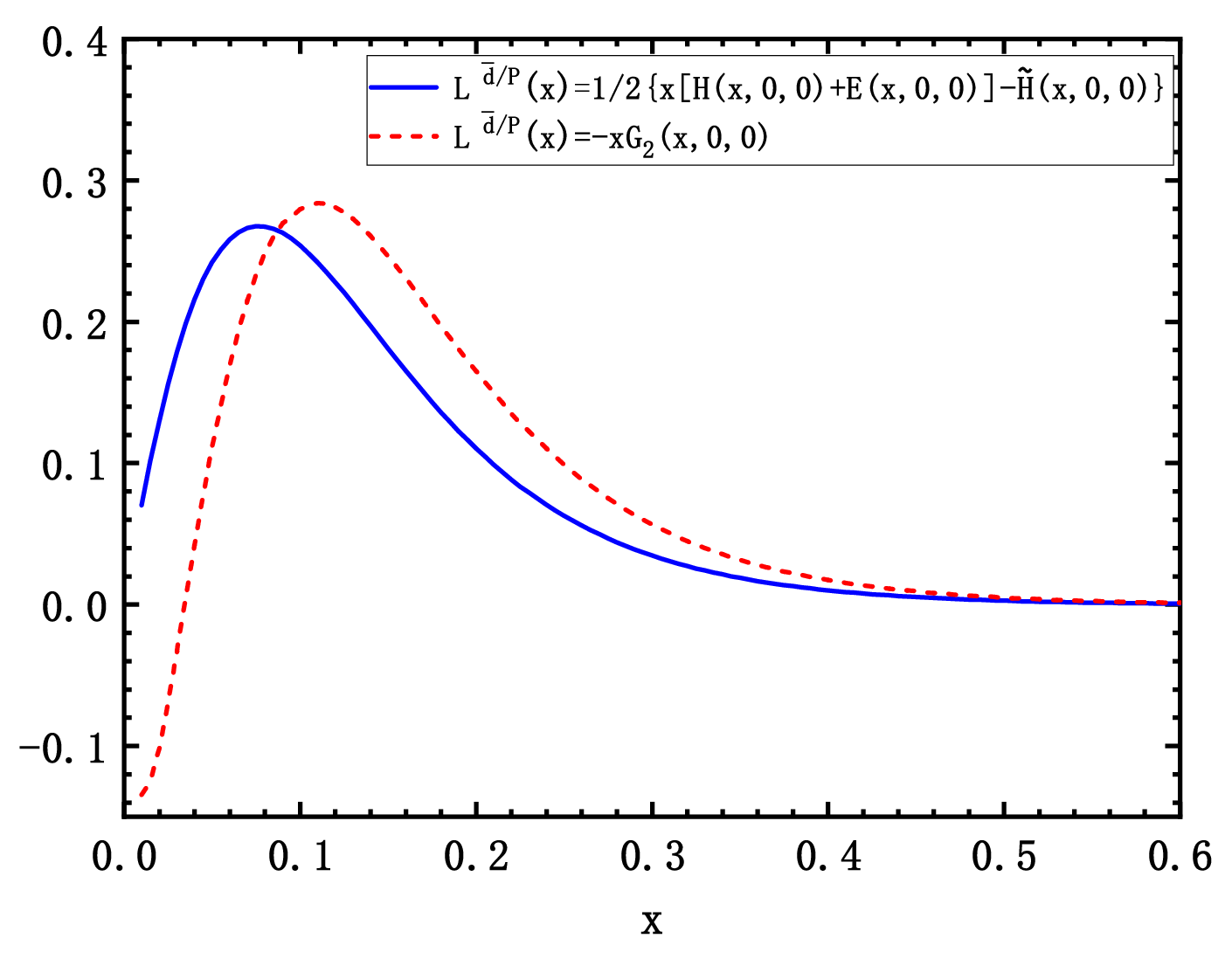}    	\end{minipage}}
     	\caption{The kinetic OAM of $\bar{u}$ (left panel) and $\bar{d}$ (right panel) quarks defined by the twist-2 and twist-3 GPDs as a function of $x$.} \label{OAM}      
     \end{figure*}
     
     \begin{figure}[htbp]
     	\centering
     	\includegraphics[width=0.9\linewidth]{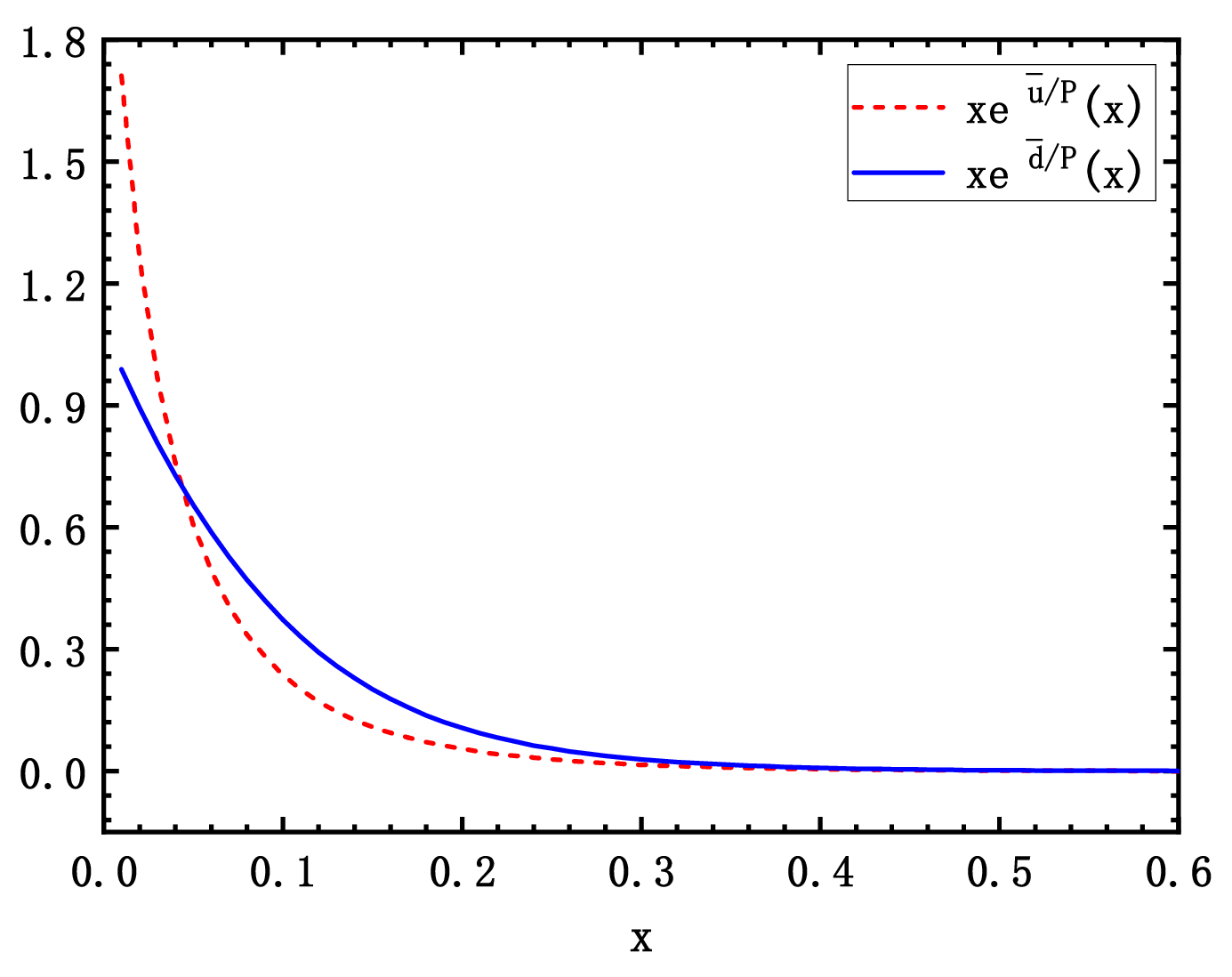}
     	\caption{The twist-3 PDFs $e^{\bar{q}/P}(x)$ (timed with $x$) of $\bar{u}$ and $\bar{d}$ quarks as a function of $x$.} 
     	\label{PDFs} 
     \end{figure}

To illustrate the $x$-dependence and $\Delta_T$-dependence of the twist-3 GPDs for the $\bar{u}$ and $\bar{d}$ quarks simultaneously, we present their 3-dimensional shapes in Figs. \ref{H2E2}-\ref{E2TpwH2Tp}. 
In the upper and lower panels of Fig. \ref{H2E2}, we plot the chiral-even GPDs associated with $\Gamma=1$, namely, $H_2^{\bar{u}/P}$, $H_2^{\bar{d}/P}$, $E_2^{\bar{u}/P}$, and $E_2^{\bar{d}/P}$ at $\xi=0$. 
We find that $xH_2^{\bar{u}/P}$ and $xH_2^{\bar{d}/P}$ are positive throughout the entire $x$ and $\Delta_T$ region. 
For fixed values of $x$, both $xH_2^{\bar{u}/P}$ and $xH_2^{\bar{d}/P}$ show a monotonically decreasing trend with increasing $\Delta_T$. 
Moreover, these distributions decrease rapidly as $x$ increases.
In contrast, we observe that the signs of $xE_2^{\bar{u}/P}$ and $xE_2^{\bar{d}/P}$ are negative  for both $\bar{u}$ and $\bar{d}$ quarks. 
The magnitudes of these distributions also decrease with increasing $x$ and $\Delta_T$, similar to the behavior of $xH_2^{\bar{u}/P}(x,0,t)$ and $xH_2^{\bar{d}/P}$. 

In Fig. \ref{wH2p}, we present the twist-3 chiral-even GPDs associated with $\Gamma=i \sigma^{+-} \gamma_{5}$, particularly $x\widetilde{H}_2^{\prime \bar{u}/P}$ and $x\widetilde{H}_2^{\prime \bar{d}/P}$ at $\xi=0$.
It is evident that both $x\widetilde{H}_2^{\prime \bar{u}/P}$ and $x\widetilde{H}_2^{\prime \bar{d}/P}$ are positive. 
Their magnitudes initially increase with $\Delta_T$ before subsequently decreasing. Additionally, $x\widetilde{H}_2^{\prime \bar{q}/P}$ of the $\bar{u}$ and $\bar{d}$ quarks peak at lower values of $x$, and the peak of these distributions shift slightly towards higher $x$ with the increase of $\Delta_T$.

In Fig.~\ref{wE2T}, we present the twist-3 chiral-odd GPDs associated with $\Gamma = \gamma^{i}$, specifically $x\widetilde{E}^{\bar{u}/P}_{2T}$ and $x\widetilde{E}^{\bar{d}/P}_{2T}$, as functions of $x$ and $\Delta_T$. 
Our findings indicate that $x\widetilde{E}^{\bar{q}/P}_{2T}$ for both $\bar{u}$ and $\bar{d}$ quarks peaks at $\Delta_T = 0$. 
Notably, for any given $\Delta_T$, the peak values of these curves increase first and then decrease with increasing $x$, and the position of the peak is found in the small $x$ region. 
Furthermore, the signs of $x\widetilde{E}^{\bar{u}/P}_{2T}$ and $x\widetilde{E}^{\bar{d}/P}_{2T}$ remain negative in the entire $x$ and $\Delta_T$ regimes.

In the upper and lower panels of Fig.~\ref{E2TpwH2Tp}, we plot the twist-3 chiral-odd GPDs corresponding to $\Gamma = \gamma^{i}\gamma_{5}$, namely, $xE^{\prime\bar{u}/P}_{2T}$, $xE^{\prime\bar{d}/P}_{2T}$ and $x\widetilde{H}^{\prime\bar{u}/P}_{2T}$, $x\widetilde{H}^{\prime\bar{d}/P}_{2T}$, as functions of $x$ and $\Delta_T$, respectively. 
It is shown that $xE^{\prime\bar{u}/P}_{2T}$ and $xE^{\prime\bar{d}/P}_{2T}$ exhibit both positive and negative values depending on the variation of $x$ and $\Delta_T$. 
Both $xE^{\prime\bar{u}/P}_{2T}$ and $xE^{\prime\bar{d}/P}_{2T}$ decrease as $\Delta_T$ increases. 
As $x$ increases, $xE^{\prime\bar{u}/P}_{2T}$ and $xE^{\prime\bar{d}/P}_{2T}$ transition from positive to negative values, indicating a sign flip in the distribution at a specific value of $x$ and $\Delta_T$. 
Similar behavior is observed for $x\widetilde{H}^{\prime\bar{u}/P}_{2T}$ and $x\widetilde{H}^{\prime\bar{d}/P}_{2T}$, which exhibit analogous 3-dimensional plots but with opposite signs compared to $xE^{\prime\bar{u}/P}_{2T}$ and $xE^{\prime\bar{d}/P}_{2T}$.
    
\subsection{Forward limit}

Sum rules for the vector and axial-vector twist-3 GPDs $G_{i}$ and $\tilde{G}_{i}$ have been derived in Ref.~\cite{Kiptily:2002nx}. 
For $x$-moments of the vector GPD $G_{2}$
\begin{align}
     	\notag&\int_{-1}^{1} d x x G_{2}(x, \xi, t) \\
     	& = \frac{1}{2} \{G_{A}(t)-\int_{-1}^{1} d xx[H(x, \xi, t)+E(x, \xi, t)]\},
\end{align}
where $G_{A}(t)$ is axial-vector form factor obtained from $x^{0}$-moments of $\widetilde{H}$.
As shown in Ref.~\cite{Ji:1996ek}, the quark angular momentum can be separated into the usual quark helicity and a gauge-invariant orbital contributions $L^q$. The kinetic OAM is related to the GPDs by Ji's sum rule:
\begin{align}
  	L^{\bar{q}} &=\int dx\frac{1}{2}\left\{x[H^{\bar{q}/P}(x,0,0)
  	+E^{\bar{q}/P}(x,0,0)]\right.\nonumber \\
  	&\left.-\widetilde{H}^{\bar{q}/P}(x,0,0)\right\}.
    \end{align}
Alternatively, the kinetic OAM can also be represented  as~\cite{Lorce:2016nxs}
\begin{align}
  	L_z^q=-\int_{-1}^1dxxG_2^q(x,0,0),
\end{align}
which is referred to as the Penttinen-Polyakov-Shuvaev-Strikman relation~\cite{Kiptily:2002nx,Penttinen:2000dg}. This implies that the relevance between twist-3 GPDs and physical observables cannot be neglected.
    
In the left and right panels of Fig. \ref{OAM}, we compare the $x$-dependence of the kinetic OAM of $\bar{u}$ and $\bar{d}$ quarks defined by the twist-2 and twist-3 GPDs, respectively. Our result reveals that the kinetic OAM defined by the twist-2 GPDs for both \(\bar{u}\) and \(\bar{d}\) quarks within the proton remains positive across the defined by the twist-2 GPDs for both $\bar{u}$ and $\bar{d}$ quarks within the proton remains positive across the entire $x$ range. While the kinetic OAM defined by the twist-3 GPDs exhibits negative values in the small $x$ region. Similar trends are observed for both cases of kinetic OAM: the peaks of these curves initially increase and then decrease as $x$ increases. Furthermore, it is noteworthy that the kinetic OAM defined by the twist-2 GPDs peaks at smaller $x$ compared to that defined by the twist-3 GPDs. Utilizing the GPDs from our model, we calculate the kinetic OAM defined by the twist-3 GPDs and obtain $L^{\bar{u}}=0.024$ and $L^{\bar{d}}=0.046$. These results are consistent with the kinetic OAM values defined by the twist-2 GPDs, as calculated in Ref. \cite{Luan:2023lmt}, where $L^{\bar{u}}=0.025$ and $L^{\bar{d}}=0.046$.
The agreement between OAM results obtained from twist-2 and twist-3 GPDs provides an important cross-check of our approach and suggests useful constraints on sea quark angular momentum contributions.
       
In the forward limit, the twist-3 GPDs $H_{2}$, $\tilde{H}_{2}^{\prime}$, $H_{2T}$ can reduce to their corresponding PDFs
\begin{align}
     		e(x)&=\operatorname*{lim}_{\Delta\to0}H_{2}(x,0,t), \\
     		h_{L}(x)&=\operatorname*{lim}_{\Delta\to0}\tilde{H}_{2}^{\prime}(x,0,t), \\
     		g_{T}(x)&=\operatorname*{lim}_{\Delta\to0}H_{2T}^{\prime}(x,0,t).
\end{align}
Among them, only has a nonzero twist-3 PDF $e(x)$. We can easily verify that our result satisfy the following relation
\begin{align}
     	e^{\bar{q}/P}(x) = \frac{m}{xM} f_{1}^{\bar{q}/P}(x), \label{e}
\end{align}
where $f_{1}^{\bar{q}/P}(x)$ has been obtained in Ref.~\cite{Luan:2022fjc}.
In Fig.~\ref{PDFs}, we depict the twist-3 PDF $xe^{\bar{q}/P}(x)$ of $\bar{u}$ and $\bar{d}$ quarks as a function of $x$. We find that the magnitudes of $xe^{\bar{u}/P}(x)$ and $xe^{\bar{d}/P}(x)$ decrease as $x$ increases. The signs of $xe^{\bar{u}/P}(x)$ and $xe^{\bar{d}/P}(x)$ are positive throughout the entire $x$ region. Particularly, the magnitude of $xe^{\bar{u}/P}(x)$ is much larger than that of $xe^{\bar{d}/P}(x)$.
    
\section{CONCLUSION}\label{Sec5}

In this work, we have investigated the twist-3 GPDs of sea quarks in the proton at zero skewness using the light-cone quark model. Our approach expresses these distributions through the overlap representation of LCWFs, providing a transparent framework for understanding sea quark contributions to nucleon structure.
To model the sea quark degrees of freedom, we considered the proton Fock state expansion that includes meson-baryon fluctuations, specifically treating the proton as a composite system of a pion and a baryon, where the pion itself consists of a quark-antiquark ($q\bar{q}$) pair. Within this framework, we derived analytic expressions for both chiral-odd and chiral-even twist-3 GPDs of $\bar{u}$ and $\bar{d}$ sea quarks.
    
Our analysis reveals two notable vanishing results within our model: the distributions $\widetilde{E}_{2}^{\bar{q}/P}(x,0,t)$ and $H_{2T}^{\prime \bar{q}/P}(x,0,t)$ are identically zero. Through careful parameter selection, we obtained numerical results for these twist-3 GPDs as functions of both the longitudinal momentum fraction $x$ and the transverse momentum transfer $\bm{\Delta}_T$.
Furthermore, we presented the numerical results for the corresponding twist-3 PDFs by taking the forward limit of the GPDs.
Particularly, we verify the relations between the twist-3 PDF $e^{\bar{q}/P}(x)$ and unpolarized PDF $f_{1}^{\bar{q}/P}(x)$ of $\bar{u}$ and $\bar{d}$ quarks
In addition, the kinetic OAMs of sea quarks defined by the twist-2 GPDs and those defined by the twist-3 GPDs are compared in our work.
The results demonstrate consistent OAM values between twist-2 and twist-3 definitions, providing important validation of our approach. 
This work offers new theoretical insights into sea quark dynamics, with implications for understanding nucleon structure and guiding future experimental measurements of higher-twist effects.

\section*{Acknowledgements}
This work is partially supported by the National Natural Science Foundation of China under grant number 12150013.

\appendix
\section{Overlap representation of twist-3 GPDs of sea quarks}
Using Eq.~(\ref{correlation}) as well as the relations in Eqs.~(\ref{relation1}) and (\ref{relation2}), the overlap representation of the twist-3 chiral-odd GPDs of sea quarks in terms of the LCWFs can be expressed as
\begin{widetext}
    \begin{align}\label{odd}
    	&\notag {H}_2\left(x,0, t\right)=\int_{x}^{1}\frac{dy}{y}\int\frac{d^2\boldsymbol{k}_T}{16\pi^3}
    \int\frac{d^2\boldsymbol{r}_T}{16\pi^3}\sum_{{\lambda_B}{\lambda_q}}\\
    \notag&\times\left[\frac{m}{2Mx}\left(\psi^{+*}_{{\lambda_B}{\lambda_q}+}
    \psi^{+}_{{\lambda_B}{\lambda_q}+}+\psi^{+*}_{{\lambda_B}{\lambda_q}-}\psi^{+}_{{\lambda_B}{\lambda_q}-}
    +\psi^{-*}_{{\lambda_B}{\lambda_q}+}\psi^{-}_{{\lambda_B}{\lambda_q}+}+\psi^{-*}_{{\lambda_B}{\lambda_q}-}
    \psi^{-}_{{\lambda_B}{\lambda_q}-}\right)\right.\displaybreak[0] \\&\left.
    +\frac{-\Delta_1+i\Delta_2}{4Mx}  \left(\psi^{+*}_{{\lambda_B}{\lambda_q}-}\psi^{+}_{{\lambda_B}{\lambda_q}+}+\psi^{-*}_{{\lambda_B}{\lambda_q}-}
    \psi^{-}_{{\lambda_B}{\lambda_q}+}\right) + \frac{\Delta_1+i\Delta_2}{4Mx}  \left(\psi^{+*}_{{\lambda_B}{\lambda_q}+}\psi^{+}_{{\lambda_B}{\lambda_q}-}+\psi^{-*}_{{\lambda_B}{\lambda_q}+}
    \psi^{-}_{{\lambda_B}{\lambda_q}-}\right) \right],\\ \notag &\frac{i\Delta_{2}}{2}{E}_2\left(x,0,t\right)=\int_{x}^{1}\frac{dy}{y}\int\frac{d^2\boldsymbol{k}_T}
    {16\pi^3}\int\frac{d^2\boldsymbol{r}_T}{16\pi^3}\sum_{{\lambda_B}{\lambda_q}}	\\\notag&\times\left[\frac{m}{2x}\left(\psi^{+*}_{{\lambda_B}{\lambda_q}+}\psi^{-}_{{\lambda_B}{\lambda_q}+}
    +\psi^{+*}_{{\lambda_B}{\lambda_q}-}\psi^{-}_{{\lambda_B}{\lambda_q}-}+\psi^{-*}_{{\lambda_B}{\lambda_q}+}
    \psi^{+}_{{\lambda_B}{\lambda_q}+}+\psi^{-*}_{{\lambda_B}{\lambda_q}-}\psi^{+}_{{\lambda_B}{\lambda_q}-}\right)
    \right.\displaybreak[0] \\&\left.+\frac{-\Delta_1+i\Delta_2}{4x}  \left(\psi^{+*}_{{\lambda_B}{\lambda_q}-}\psi^{-}_{{\lambda_B}{\lambda_q}+}+\psi^{-*}_{{\lambda_B}{\lambda_q}-}
    \psi^{+}_{{\lambda_B}{\lambda_q}+}\right) + \frac{\Delta_1+i\Delta_2}{4x}  \left(\psi^{+*}_{{\lambda_B}{\lambda_q}+}\psi^{-}_{{\lambda_B}{\lambda_q}-}+\psi^{-*}_{{\lambda_B}{\lambda_q}+}
    \psi^{+}_{{\lambda_B}{\lambda_q}-}\right) \right],\\
    \notag &\frac{-\Delta_{1}}{2}\widetilde{E}_2\left(x,0,t\right)=\int_{x}^{1}\frac{dy}{y}
    \int\frac{d^2\boldsymbol{k}_T}{16\pi^3}\int\frac{d^2\boldsymbol{r}_T}{16\pi^3}\sum_{{\lambda_B}{\lambda_q}}	\\ \notag&\times\left[\frac{\Delta_{1}-i\Delta_{2}}{4x}\left(\psi^{+*}_{{\lambda_B}{\lambda_q}-}
    \psi^{-}_{{\lambda_B}{\lambda_q}+}+\psi^{-*}_{{\lambda_B}{\lambda_q}-}\psi^{+}_{{\lambda_B}{\lambda_q}+}\right)
    +\frac{\Delta_{1}+i\Delta_{2}}{4x}\left(\psi^{+*}_{{\lambda_B}{\lambda_q}+}\psi^{-}_{{\lambda_B}{\lambda_q}-}
    +\psi^{-*}_{{\lambda_B}{\lambda_q}+}\psi^{+}_{{\lambda_B}{\lambda_q}-}\right) \right],\\
    \notag &\widetilde{H}_{2}^{\prime}\left(x,0,t\right)=-\int_{x}^{1}\frac{dy}{y}\int\frac{d^2\boldsymbol{k}_T}
    {16\pi^3}\int\frac{d^2\boldsymbol{r}_T}{16\pi^3}\sum_{{\lambda_B}{\lambda_q}} 	\\
    \notag&\times\left[\frac{m}{2Mx}\left(\psi^{+*}_{{\lambda_B}{\lambda_q}-}\psi^{+}_{{\lambda_B}{\lambda_q}-}
    -\psi^{-*}_{{\lambda_B}{\lambda_q}-}\psi^{-}_{{\lambda_B}{\lambda_q}-}-\psi^{+*}_{{\lambda_B}{\lambda_q}+}
    \psi^{+}_{{\lambda_B}{\lambda_q}+}+\psi^{-*}_{{\lambda_B}{\lambda_q}+}\psi^{-}_{{\lambda_B}{\lambda_q}+}\right)
    \right.\displaybreak[0] \\&\left.+\frac{-k_1+ik_2}{2Mx}  \left(\psi^{+*}_{{\lambda_B}{\lambda_q}-}
    \psi^{+}_{{\lambda_B}{\lambda_q}+}-\psi^{-*}_{{\lambda_B}{\lambda_q}-}\psi^{-}_{{\lambda_B}{\lambda_q}+}\right) + \frac{-(k_1+ik_2)}{2Mx} \left(\psi^{+*}_{{\lambda_B}{\lambda_q}+}\psi^{+}_{{\lambda_B}{\lambda_q}-}-
    \psi^{-*}_{{\lambda_B}{\lambda_q}+}\psi^{-}_{{\lambda_B}{\lambda_q}-}\right) \right].
\end{align}
    
Similarly, the chiral-even twist-3 GPDs in the overlap representation can be expressed as
\begin{align}\label{even}
    \notag &\frac{i\boldsymbol{\Delta}_T^{2}}{2}\widetilde{E}_{2T}\left(x,0, t\right)=-\int_{x}^{1}\frac{dy}{y}\int\frac{d^2\boldsymbol{k}_T}{16\pi^3}
    \int\frac{d^2\boldsymbol{r}_T}{16\pi^3}\sum_{{\lambda_B}{\lambda_q}}		\\
    \notag&\times\left\{\Delta_{2}\left[\frac{2k_1+i\Delta_{2}}{4x}\left(\psi^{+*}_{{\lambda_B}{\lambda_q}-}
    \psi^{+}_{{\lambda_B}{\lambda_q}-}-\psi^{-*}_{{\lambda_B}{\lambda_q}-}\psi^{-}_{{\lambda_B}{\lambda_q}-}\right)
    +\frac{2k_1-i\Delta_{2}}{4x}\left(\psi^{+*}_{{\lambda_B}{\lambda_q}+}\psi^{+}_{{\lambda_B}{\lambda_q}+}-
    \psi^{-*}_{{\lambda_B}{\lambda_q}+}\psi^{-}_{{\lambda_B}{\lambda_q}+}\right) \right]\right.\\&\left.		-\Delta_{1}\left[\frac{2k_2-i\Delta_{1}}{4x}\left(\psi^{+*}_{{\lambda_B}{\lambda_q}-}
    \psi^{+}_{{\lambda_B}{\lambda_q}-}-\psi^{-*}_{{\lambda_B}{\lambda_q}-}\psi^{-}_{{\lambda_B}{\lambda_q}-}\right)
    +\frac{2k_2+i\Delta_{1}}{4x}\left(\psi^{+*}_{{\lambda_B}{\lambda_q}+}\psi^{+}_{{\lambda_B}{\lambda_q}+}
    -\psi^{-*}_{{\lambda_B}{\lambda_q}+}\psi^{-}_{{\lambda_B}{\lambda_q}+}\right) \right]	\right\},\\\notag &\frac{i\boldsymbol{\Delta}_T^{2}}{2}({E}_{2T}^{\prime}+2\widetilde{H}_{2T}^{\prime})
    =\int_{x}^{1}\frac{dy}{y}\int\frac{d^2\boldsymbol{k}_T}{16\pi^3}\int\frac{d^2\boldsymbol{r}_T}{16\pi^3}
    \sum_{{\lambda_B}{\lambda_q}}\\
    \notag&\times\left\{\Delta_2\left[\frac{m}{2x}\left(\psi^{+*}_{{\lambda_B}{\lambda_q}-}
    \psi^{+}_{{\lambda_B}{\lambda_q}+}+\psi^{-*}_{{\lambda_B}{\lambda_q}-}\psi^{-}_{{\lambda_B}{\lambda_q}+}
    +\psi^{+*}_{{\lambda_B}{\lambda_q}+}\psi^{+}_{{\lambda_B}{\lambda_q}-}+\psi^{-*}_{{\lambda_B}{\lambda_q}+}
    \psi^{-}_{{\lambda_B}{\lambda_q}-}\right)\right.\right.\displaybreak[0] \\&\left.\left.
    		+\frac{-2k_1+i\Delta_2}{4x}  \left(\psi^{+*}_{{\lambda_B}{\lambda_q}+}\psi^{+}_{{\lambda_B}{\lambda_q}+}+\psi^{-*}_{{\lambda_B}{\lambda_q}+}
    \psi^{-}_{{\lambda_B}{\lambda_q}+}\right) + \frac{2k_1+i\Delta_2}{4x}  \left(\psi^{+*}_{{\lambda_B}{\lambda_q}-}\psi^{+}_{{\lambda_B}{\lambda_q}-}+\psi^{-*}_{{\lambda_B}{\lambda_q}-}
    \psi^{-}_{{\lambda_B}{\lambda_q}-}\right) \right]\right. \\&\left.		-\Delta_1\left[\frac{im}{2x}\left(-\psi^{+*}_{{\lambda_B}{\lambda_q}-}\psi^{+}_{{\lambda_B}{\lambda_q}+}
    -\psi^{-*}_{{\lambda_B}{\lambda_q}-}\psi^{-}_{{\lambda_B}{\lambda_q}+}+\psi^{+*}_{{\lambda_B}{\lambda_q}+}
    \psi^{+}_{{\lambda_B}{\lambda_q}-}+\psi^{-*}_{{\lambda_B}{\lambda_q}+}\psi^{-}_{{\lambda_B}{\lambda_q}-}\right)
    \right.\right.\displaybreak[0] \\&\left.\left.+\frac{-2k_2-i\Delta_1}{4x}  \left(\psi^{+*}_{{\lambda_B}{\lambda_q}+}\psi^{+}_{{\lambda_B}{\lambda_q}+}+\psi^{-*}_{{\lambda_B}{\lambda_q}+}
    \psi^{-}_{{\lambda_B}{\lambda_q}+}\right) + \frac{2k_2-i\Delta_1}{4x}  \left(\psi^{+*}_{{\lambda_B}{\lambda_q}-}\psi^{+}_{{\lambda_B}{\lambda_q}-}+\psi^{-*}_{{\lambda_B}{\lambda_q}-}
    \psi^{-}_{{\lambda_B}{\lambda_q}-}\right) \right]\right\},\\
    \notag &{H}_{2T}^{\prime}+\frac{\boldsymbol{\Delta}_T^{2}}{4M^{2}}\widetilde{H}_{2T}^{\prime}=\int_{x}^{1}\frac{dy}{y}
    \int\frac{d^2\boldsymbol{k}_T}{16\pi^3}\int\frac{d^2\boldsymbol{r}_T}{16\pi^3}\sum_{{\lambda_B}{\lambda_q}}\\
    \notag&\times\left\{\left[\frac{m}{4Mx}\left(\psi^{+*}_{{\lambda_B}{\lambda_q}-}\psi^{-}_{{\lambda_B}{\lambda_q}+}
    +\psi^{-*}_{{\lambda_B}{\lambda_q}-}\psi^{+}_{{\lambda_B}{\lambda_q}+}+\psi^{+*}_{{\lambda_B}{\lambda_q}+}
    \psi^{-}_{{\lambda_B}{\lambda_q}-}+\psi^{-*}_{{\lambda_B}{\lambda_q}+}\psi^{+}_{{\lambda_B}{\lambda_q}-}\right)
    \right.\right.\displaybreak[0] \\&\left.\left.+\frac{-2k_1+i\Delta_2}{8Mx}  \left(\psi^{+*}_{{\lambda_B}{\lambda_q}+}\psi^{-}_{{\lambda_B}{\lambda_q}+}+\psi^{-*}_{{\lambda_B}{\lambda_q}+}
    \psi^{+}_{{\lambda_B}{\lambda_q}+}\right) + \frac{2k_1+i\Delta_2}{8Mx}  \left(\psi^{+*}_{{\lambda_B}{\lambda_q}-}\psi^{-}_{{\lambda_B}{\lambda_q}-}+\psi^{-*}_{{\lambda_B}{\lambda_q}-}
    \psi^{+}_{{\lambda_B}{\lambda_q}-}\right) \right]\right. \\&\left.		+i\left[\frac{im}{4Mx}\left(-\psi^{+*}_{{\lambda_B}{\lambda_q}-}\psi^{-}_{{\lambda_B}{\lambda_q}+}
    +\psi^{-*}_{{\lambda_B}{\lambda_q}-}\psi^{+}_{{\lambda_B}{\lambda_q}+}+\psi^{+*}_{{\lambda_B}{\lambda_q}+}
    \psi^{-}_{{\lambda_B}{\lambda_q}-}-\psi^{-*}_{{\lambda_B}{\lambda_q}+}\psi^{+}_{{\lambda_B}{\lambda_q}-}\right)
    \right.\right.\displaybreak[0] \\&\left.\left.+\frac{-2k_2-i\Delta_1}{8Mx}  \left(\psi^{+*}_{{\lambda_B}{\lambda_q}+}\psi^{-}_{{\lambda_B}{\lambda_q}+}-\psi^{-*}_{{\lambda_B}{\lambda_q}+}
    \psi^{+}_{{\lambda_B}{\lambda_q}+}\right) + \frac{2k_2-i\Delta_1}{8Mx}  \left(\psi^{+*}_{{\lambda_B}{\lambda_q}-}\psi^{-}_{{\lambda_B}{\lambda_q}-}-\psi^{-*}_{{\lambda_B}{\lambda_q}-}
    \psi^{+}_{{\lambda_B}{\lambda_q}-}\right) \right]\right\},\\
    \notag &\frac{i\Delta_{1}\Delta_{2}}{M}\widetilde{H}_{2T}^{\prime}\left(x,0,t\right)
    =\int_{x}^{1}\frac{dy}{y}\int\frac{d^2\boldsymbol{k}_T}{16\pi^3}\int\frac{d^2\boldsymbol{r}_T}{16\pi^3}
    \sum_{{\lambda_B}{\lambda_q}}\\
    \notag &\times\left\{\left[\frac{m}{2x}\left(\psi^{+*}_{{\lambda_B}{\lambda_q}-}
    \psi^{-}_{{\lambda_B}{\lambda_q}+}-\psi^{-*}_{{\lambda_B}{\lambda_q}-}\psi^{+}_{{\lambda_B}{\lambda_q}+}
    +\psi^{+*}_{{\lambda_B}{\lambda_q}+}\psi^{-}_{{\lambda_B}{\lambda_q}-}-\psi^{-*}_{{\lambda_B}{\lambda_q}+}
    \psi^{+}_{{\lambda_B}{\lambda_q}-}\right)\right.\right.\displaybreak[0] \\&\left.\left. +\frac{-2k_1+i\Delta_2}{4x}\left(\psi^{+*}_{{\lambda_B}{\lambda_q}+}\psi^{-}_{{\lambda_B}{\lambda_q}+}
    -\psi^{-*}_{{\lambda_B}{\lambda_q}+}\psi^{+}_{{\lambda_B}{\lambda_q}+}\right) + \frac{2k_1+i\Delta_2}{4x}  \left(\psi^{+*}_{{\lambda_B}{\lambda_q}-}\psi^{-}_{{\lambda_B}{\lambda_q}-}-\psi^{-*}_{{\lambda_B}{\lambda_q}-}
    \psi^{+}_{{\lambda_B}{\lambda_q}-}\right) \right]\right. \\&\left.		-i\left[\frac{im}{2x}\left(-\psi^{+*}_{{\lambda_B}{\lambda_q}-}\psi^{-}_{{\lambda_B}{\lambda_q}+}
    -\psi^{-*}_{{\lambda_B}{\lambda_q}-}\psi^{+}_{{\lambda_B}{\lambda_q}+}+\psi^{+*}_{{\lambda_B}{\lambda_q}+}
    \psi^{-}_{{\lambda_B}{\lambda_q}-}+\psi^{-*}_{{\lambda_B}{\lambda_q}+}\psi^{+}_{{\lambda_B}{\lambda_q}-}\right)
    \right.\right.\displaybreak[0] \\&\left.\left.+\frac{-2k_2-i\Delta_1}{4x}  \left(\psi^{+*}_{{\lambda_B}{\lambda_q}+}\psi^{-}_{{\lambda_B}{\lambda_q}+}+\psi^{-*}_{{\lambda_B}{\lambda_q}+}
    \psi^{+}_{{\lambda_B}{\lambda_q}+}\right) + \frac{2k_2-i\Delta_1}{4x}  \left(\psi^{+*}_{{\lambda_B}{\lambda_q}-}\psi^{-}_{{\lambda_B}{\lambda_q}-}+\psi^{-*}_{{\lambda_B}{\lambda_q}-}
    \psi^{+}_{{\lambda_B}{\lambda_q}-}\right) \right]\right\}.\label{even}
\end{align}	
\end{widetext}
    
For simplicity, here we use $\psi^{\Lambda^{\prime}*}_{{\lambda_B}{\lambda_q}{\lambda_{\bar{q}}^{\prime}}}$ and $\psi^{\Lambda}_{{\lambda_B}{\lambda_q}{\lambda_{\bar{q}}}}$ to represent  $\psi^{\Lambda^{\prime}*}_{{\lambda_B}{\lambda_q}{\lambda_{\bar{q}}^{\prime}}}
(x,y,\boldsymbol{k}^{\prime\prime}_T,\boldsymbol{r}^{\prime\prime}_T)$ and  $\psi^{\Lambda}_{{\lambda_B}{\lambda_q}{\lambda_{\bar{q}}}}(x,y,\boldsymbol{k}^{\prime}_T,\boldsymbol{r}^{\prime}_T)$, respectively. And 
\begin{align}	
\nonumber\boldsymbol{k}^{{\prime}{\prime}}_T&=\boldsymbol{k}_T-\frac{1}{2}(1-x)\boldsymbol{\Delta}_T  \\
\boldsymbol{k}^{\prime}_T&=\boldsymbol{k}_T+\frac{1}{2}(1-x)\boldsymbol{\Delta}_T,
\end{align}	
are the transverse momenta for the final-state and initial-state struck antiquarks,       
\begin{align}	
\nonumber -\boldsymbol{r}^{{\prime}{\prime}}_T&=-\boldsymbol{r}_T+\frac{1}{2}(1-y)\boldsymbol{\Delta}_T   \\ \nonumber-\boldsymbol{r}^{\prime}_T&=-\boldsymbol{r}_T-\frac{1}{2}(1-y)\boldsymbol{\Delta}_T\\
\nonumber(\boldsymbol{r}_T-\boldsymbol{k}_T)^{{\prime}{\prime}}&=
(\boldsymbol{r}_T-\boldsymbol{k}_T)+\frac{1}{2}(y-x)\boldsymbol{\Delta}_T \\
(\boldsymbol{r}_T-\boldsymbol{k}_T)^{\prime}&=	(\boldsymbol{r}_T-\boldsymbol{k}_T)-\frac{1}{2}(y-x)\boldsymbol{\Delta}_T,
\end{align}	
are the transverse momenta for the final and initial spectators $B$ and $q$, respectively.
 	
\end{document}